\documentclass[iop]{emulateapj}
\usepackage{graphicx}
\usepackage{hyperref}
\usepackage{longtable}
\usepackage{ulem}
\usepackage[dvipsnames]{xcolor}

\shorttitle{HD~141569}
\shortauthors{Jensen et al.}

\begin{document}

\title{Modeling of CO ro-vibrational line emission of HD~141569}

\author{Stanley K. Jensen Jr.}
\affil{Department of Physics and Astronomy, 118 Kinard Laboratory, Clemson University, Clemson, SC 29634-0978}

\author{Sean D. Brittain}
\affil{Department of Physics and Astronomy, 118 Kinard Laboratory, Clemson University, Clemson, SC 29634-0978}

\author{Joan R. Najita}
\affil{NSF's NOIRLab, 950 North Cherry Avenue, Tucson, AZ 85719}

 \author{John S. Carr}
\affil{Department of Astronomy, University of Maryland, College Park, MD 20742, USA}

\begin{abstract}
HD~141569 is a Herbig Ae/Be star that straddles the boundary between the transition disks and debris disks. It is a low dust mass disk that reveals numerous structural elements (e.g. gaps and rings) that may point to young planets. It also exhibits a reservoir of CO gas observed at both millimeter and IR wavelengths. Previous observations \citep{goto06} reported a possible asymmetry in the CO gas emission. Herein the IR ro-vibrational emission lines are analyzed and modeled both spectroscopically and spectroastrometrically. We find emission features from both $^{12}$CO and $^{13}$CO isotopologues heated to a temperature of approximately 200~K in the radial extent of 13 to 60 au. We do not see evidence of the previously reported asymmetry in CO emission, our results being consistent with a Keplerian, axisymmetric emitting region. This raises the question of whether the emission profile may be evolving in time, possibly as a result of an orbiting feature in the inner disk such as a planet.
\end{abstract}

\section{Introduction}
During the early stages of formation, stars are frequently surrounded by protoplanetary disks. These disk are shaped by a variety of factors, such as their host stars or planets that may form within them. 
Planets in the process of forming can affect the disk in potentially dramatic and observable ways, such as inducing gaps, spiral arms, and disk eccentricities (e.g. \citealt{kley06,dong15spiral,dong15gap}). Observations of the gas and dust within these disks can give insight into the kind of processes at work in this early stage of stellar and planetary development.

Transition disks in particular are a subclass of interest. While protoplanetary disks create an IR excess, resulting from emission of the dust, transition disks lack a NIR excess. This is understood as a gap or cavity in the extent of the dust near the central star. It may also be the case, though it is not strictly necessary, that gas also is lacking in these regions. The cause of such observed gaps is the subject of ongoing research, both generally and for specific systems, with the most exciting explanation being a planet or planets in the act of formation (e.g. \citealt{strom89,najita07,najita15,espaillat14,vandermarel16}).

Definitively detecting planets within disks has proven difficult, 
although recently a companion of PDS 70 was detected by \cite{keppler18} using the VLT instrument SPHERE. In general, direct imaging to identify high-mass planets and their circumplanetary disk (CPD) or envelope has been less successful than expected. As a result, there is a need for alternatives such as potential signposts that are more readily observable. Possible alternatives are indirect signatures of planet-disk interaction, such as the aforementioned gaps/holes in disks or spiral structure. However, these signposts can be generated by other mechanisms (e.g. photoevaporation and gravitational instability, respectively), so the direct detection of a massive companion is crucial in order to validate these signposts.

A complementary approach to detecting forming planets is through the line emission formed in their circimplanetary disks. Circumplanetary disks are intrinsically three-dimensional structures with complicated flow patterns \citep[e.g.,][]{tanigawa2012, Ayliffe2012, Gressel2013b, Szulagyi2017}. Simulations indicate the size of the disk can be anywhere from one-third to the full extent of the Hill sphere \cite[][]{quillen2004, Ayliffe2009-acc, Ayliffe2009-disc, martin2011cycle, szulagyi2014-acc, szulagyi2016-disc} with temperatures $\gtrsim 1000$~K in steady state \citep{szulagyi2017-temp} and perhaps much higher during an outburst \citep{Zhu2015b}. At these sizes and temperatures, ro-vibrational emission of CO in the NIR is detectable \citep[e.g.,][]{najita2003}. An example of the potential of this method comes from HD~100546 \citep{brittain19},
where 
we have inferred the presence of the inner companion from a component of the $v=1-0$ ro-vibrational CO emission that varies relative to the stable hot band (i.e., $v^\prime \geq 2$, where $v^\prime$ is the upper vibrational state)  emission. The variation of the 
observed 
Doppler shift and the spectroastrometric signal can be explained by an emission source in a Keplerian orbit near the edge of the disk gap. Here we present a similar study of the ro-vibrational CO emission emerging from HD~141569.

HD~141569 is a 5.9 Myr old \citep{wichittanakom2020} Herbig Ae/Be star 110.6$\pm$0.54 parsecs away \citep{gaiamission,gaiadr2}. A low-mass transition disk on the cusp of becoming a debris disk, the disk in the system is a few hundred au in total
extent
\citep{weinberger99,wyatt15}. Observations in the (sub)millimeter show the gas extending to $\sim$270 au through rotational $^{12}$CO lines and dust to nearly 400 au using continuum emission 
\citep{difolco20}. It also shows signs of belonging to the $\lambda$ Bo\"{o} group of stars in its photospheric abundances \citep{gray98}. This group is marked by a depletion in refractory elements, indicating accretion of the surrounding disk material, which has undergone some degree of gas-dust separation, onto the star (e.g. \citealt{king94,murphy15}).

Various structural aspects have been observed, some of which show asymmetries. \cite{marsh2002} noted a decrease in optical depth inside of 30 au based on mid-infrared imaging. The SED of HD~141569 indicates that the inner 10 au has been cleared \citep{malfait98}. Tightly wound, large-scale spirals with radii of a few hundred au were observed with the ACS on HST by \cite{clampin03}, with both planets and tidal interactions from HD~141569's stellar companions being put forth as possible causes. More recently \cite{perrot16} observed ring-like structures within 40 and 100 au using SPHERE. Observations with the STIS instrument on HST by \cite{konishi16} detected an additional component of the disk inward of, and separated by a gap from, those seen by \cite{clampin03}. This component partially overlaps the inner region discussed by \cite{marsh2002} and shows spiral structure at $\sim$120 au.

CO is one of the most abundant molecules in protoplanetary disks. 
While millimeter emission lines result 
from pure rotational transitions of CO which 
are sensitive to the cool gas, 
observations of 
the near infrared lines from ro-vibrational CO transitions are sensitive to the warm gas in the inner disk regions. Within the disk, the rotational population is determined by thermal excitation, while the vibration population is due to UV fluorescence from stellar irradiation (e.g. \citealt{brittain07}). These infrared emission features can provide useful information as to the distribution and physical properties of the gas within the disk. Previous studies of HD~141569 have observed the infrared CO ro-vibration emission to investigate the physical conditions and extent of the disk, such as inner/outer radii and temperature (e.g. \citealt{brittain02,brittain03,goto06,brittain07,thi14,vanderplas15}).

More recent ALMA observations by \cite{flaherty16} using CO(3-2) and CO(1-0) emission determined an inner radius of 29$^{+14}_{-20}$ au. Imaging by \cite{miley18} of the midplane with ALMA using $^{13}$CO (2-1) emission found the gas to be asymmetrically distributed. They also found a midplane counterpart to a ring previously discovered in scattered light \citep{biller15} at $\sim$220 au, and found the CO gas, particularly $^{13}$CO, extends to approximately the same region. Thus the size of the gas disk is smaller than the radial extent of the dust, which extends to $\sim$400 au \citep{clampin03}. This is highly unusual as the dust component of most disks is smaller than the gas disk. This is generally understood to be the result of the inward migration of dust due to gas drag. This population of dust that extends beyond the gas may indicate that it is second generation dust in keeping with the suggestion that HD~141569 bridges the transition from transition disk to debris disk.

The derived inclination of the disk of HD~141569 has varied somewhat with different observations. Near-infrared observations with the Hubble Space Telescope (HST) instrument NICMOS by \cite{weinberger99} found an inclination of $51^{\circ}\pm 3^{\circ}$. Optical images taken with HST instrument STIS by \cite{mouillet01} found $55^{\circ}\pm 1.2^{\circ}$. Observations using sub-millimeter facilities also found differing inclinations of $60^{\circ}\pm 3^{\circ}$ (Submillimeter Array, CARMA; \citealt{flaherty16}) and $53.4^{\circ}$$^{+1.2^{\circ}}_{-0.9^{\circ}}$ (ALMA; \citealt{white16}).

Investigations of the double-peaked CO ro-vibration lines in the disk of HD~141569 have shown a region deficient or entirely devoid of gas inside of 9-15 au \citep{goto06,brittain07} and the CO emission likely results from a population of gas beyond 17 au with a rotational temperature $\sim$200~K \citep{brittain02,brittain03}. \cite{thi14} found the gas of the inner region of the disk to be in the range of 100 to 300~K. This is cooler than 
is found for the near-infrared CO emission from 
many other transition disks such as HD~100546 ($\sim$1000 K, \citealt{brittain09}), HD~97058 ($\sim$1300, \citealt{vanderplas15}), and HD~179218 (600-700 K, \citealt{brittain18}). An example of a disk with a more similar rotational temperature is Oph~IRS~48 at 260~K \citep{brown12}.

A possible north-south asymmetry is observed in the spatially resolved ro-vibrational spectrum of CO. 
\cite{goto06}
reported an inner hole size of 11$\pm$2 au using a distance of 108 pc based on \cite{merin04}. Correcting for the current Gaia distance of 110 pc makes a marginal difference to the cavity radius ($\sim$11.25 au). Such asymmetries may be caused by, for instance, giant planets which induce vortices \citep{hammer17} or by an eccentric disk induced by a massive planet \citep{kley06}. These features may be useful in the discovery of planets as a way to differentiate from a transition disk cavity that is created by photoevaporation or grain growth. In the case of an eccentric disk induced by a massive planet, the orientation will precess very slowly (~1$^{\circ}$ per 100 orbits, \citealt{kley06}), and thus a resulting line asymmetry will be observable over decadal timescales.

In this paper we present high resolution ($\lambda$/$\Delta \lambda \sim$ 75,000) M-band spectroscopy of HD~141569. We also measure the spectroastrometric signal of the emission lines. Spectroastrometry is a technique for determining spatial information via spectroscopic observations originally described by \cite{beckers82} and \cite{christy83}. Specifically it involves the measurement of the centroid of the PSF of the spectrum on the detector as a way of determining the center of light (in analogy to, for instance, center of mass) of the observed object. 
The presence of 
a close companion or rotating disk around a point-like star will shift the centroid away from the position of the central object in a characteristic pattern as a function of velocity.
On longer timescales 
and 
with multiple observations, this 
technique 
can be used to identify variable features, including potential (proto-) planetary companions \citep{brittain14,brittain15,brittain19} or disk winds \citep{whelan08,pontoppidan08,pontoppidan11}. We compare the CO line profiles and CO temperature to previous observations, and use the spectroastrometric signal to test whether the emission is consistent with gas arising in Keplerian orbit.

\section{Observations and Reductions}
The data were obtained on April 19, 2017 between 11:59 and 16:05 UT with the iSHELL cross-dispersion echelle spectrograph at the Infrared Telescope Facility (IRTF) \citep{rayner16}. The date was chosen to ensure that the target CO emission lines would be shifted out of corresponding telluric absorption features. Observations used the M2 mode (16 orders with wavelength range 4.52-5.25 $\mu$m; roughly 1910-2220 cm$^{-1}$) with an ABBA nodding pattern which allowed us to combine the images in an A-B-B+A pattern so as to remove sky emission to first order. Seeing for the night was approximately 0$^{\prime \prime}$.8-0$^{\prime \prime}$.9. Targets included the science star HD~141569 in three sets, each of which was further separated into position angles (PAs) of 356$^{\circ}$ and 176$^{\circ}$ along the semi-major axis of the disk, and additional standard stars HR~6556 and HR~5793, the former standard associated with the first science set and the latter for the other two. Observations for HD~141569 were taken in 15 second exposures with four coadds at two position angles, 356$^{\circ}$ and 176$^{\circ}$, in each of the three sets for a total integration time of 180 minutes. The two position angles are necessary in order to correct any instrumental artifacts that may affect the spectroastrometric centroid measurement. Both standards were also observed with 15 second exposures and 4 coadds. Flats were taken after observations were completed for each target. The slit width in all cases was 0$^{\prime \prime}$.375 providing a spectral resolution of $\sim$75,000. Information regarding the details of the observations are summarized in Table \ref{tab:obstab1}.

The obtained data were reduced using a standard procedure, described in \cite{brittain18}. General reductions were carried out order-by-order using custom IDL code. The ABBA nodding in addition to removing sky emission also resulted in an A and B beam spectrum in each order. The A and B beams were individually fit to polynomials to rectify the order. After rectification, dispersion correction was applied using the Spectrum Synthesis Program (SSP) telluric atmospheric radiance model \citep{kunde74}. Then the spectrum was extracted from a rectangular region by selecting high-signal rows of each beam and summing columns into new single-row elements to form each beam's spectrum. Following this, the resulting A and B beams were combined. After reductions were complete for both the science target and the telluric standard to this point, the ratio of the two was taken to remove telluric features from the science target.

The wavelength calibration was refined after combining beams (using SSP) and after ratioing (using the standard) as necessary. In the event that any large scale variation in the continuum remained after telluric correction, a low-order polynomial was used to flatten the continuum.

In order to compare the fully reduced data with the model (described in section \ref{modeling}), which generates flux in physical units erg/s/cm$^2$/cm$^{-1}$, the data continuum level must be properly converted from the relative flux of the science-to-standard ratio. This requires knowledge of the continuum level of the star in the relevant wavelength range. For the case here, 4.6 $\mu$m ALLWISE photometry \citep{allwise,cutri_allwise2} was used, obtained through the VizieR database \citep{ochsenbein00}, yielding a continuum value of 1.86$\times 10^{-13}$ erg/s/cm$^2$/cm$^{-1}$.

\begin{figure}[htb!]
\includegraphics[trim=50 50 50 50,width=0.5\textwidth]{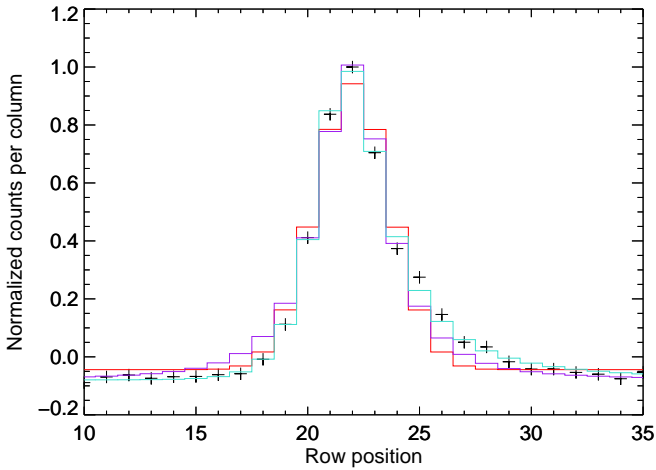}
\caption{\label{fig:smplPSF} Sample PSF (HD~141569 set 2, PA 356$^{\circ}$, order 108, A beam) with three different function fits. The normalized, column-averaged counts are shown as black plus-signs. Gaussian fit in red, Moffat fit in purple, and skewed Moffat fit in turquoise. The Gaussian and Moffat fits were generated by \texttt{MPFITPEAK} while the final skewed Moffat fit was found with \texttt{MPFITEXPR}.}
\end{figure}

Centroid measurements were performed using the rectified orders generated before spectral extraction with \texttt{MPFITEXPR}, a part of the IDL code \texttt{MPFIT} \citep{markwardt09}, iteratively for each column of each beam. Here, a skewed Moffat function was used, of the form below:
\begin{equation}
    f = p_0 \left( \frac{(x-p_1)^2}{u^2} + 1 \right) ^{-p_4} + p_5
    \label{eq:mof_skew_f}
\end{equation}
\begin{equation}
    u = \frac{2\cdot p_2}{1+e^{p_3 (x-p_1)}}
    \label{eq:mof_skew_u}
\end{equation}
Where $p_0$ is the amplitude, $p_1$ is the center, $p_2$ is the half-width at half-maximum (HWHM), $p_3$ is the skew parameter, $p_4$ is controls with broadness of the wings, and $p_5$ is the zero-level of the function. The skewed Moffat function was used because a Gaussian was determined to not be as accurate a description of the shape of the point spread function (PSF; see Fig. \ref{fig:smplPSF} for comparison).
An average PSF was determined using the regions of the continuum identified during the reduction process. The PSF was used to determine a 
``typical''
fit as a basis for the fitting of the individual columns. Once the typical fit was found, all parameters were fixed except for those pertaining to the height and centroid position. The PSF for each column was then fit across the entirety of each beam. This was repeated for each position angle and order. An additional attempt was made to look for variation in the width of the PSF across the continnum and the emission features. However, no consistent variation across the spectrum was found. Any underlying variation requires higher signal-to-noise and resolution than 
were 
obtained here.

Artifacts that affect the measurement of the centroid of the spectrum may result from the instrument itself and must be corrected. This is accomplished by the rotation of the slit position during observations. Since the observed position angles have a 180$^{\circ}$ difference, the spectroastrometric signal is inverted in one compared to the other while any instrumental artifacts should remain unchanged. Thus averaging one position angle with the negative of the other removes these artifacts. (e.g. \citealt{whelan08,pontoppidan08})

\begin{figure}[htb!]
\includegraphics[trim=0 25 0 25, width=1.0\textwidth]{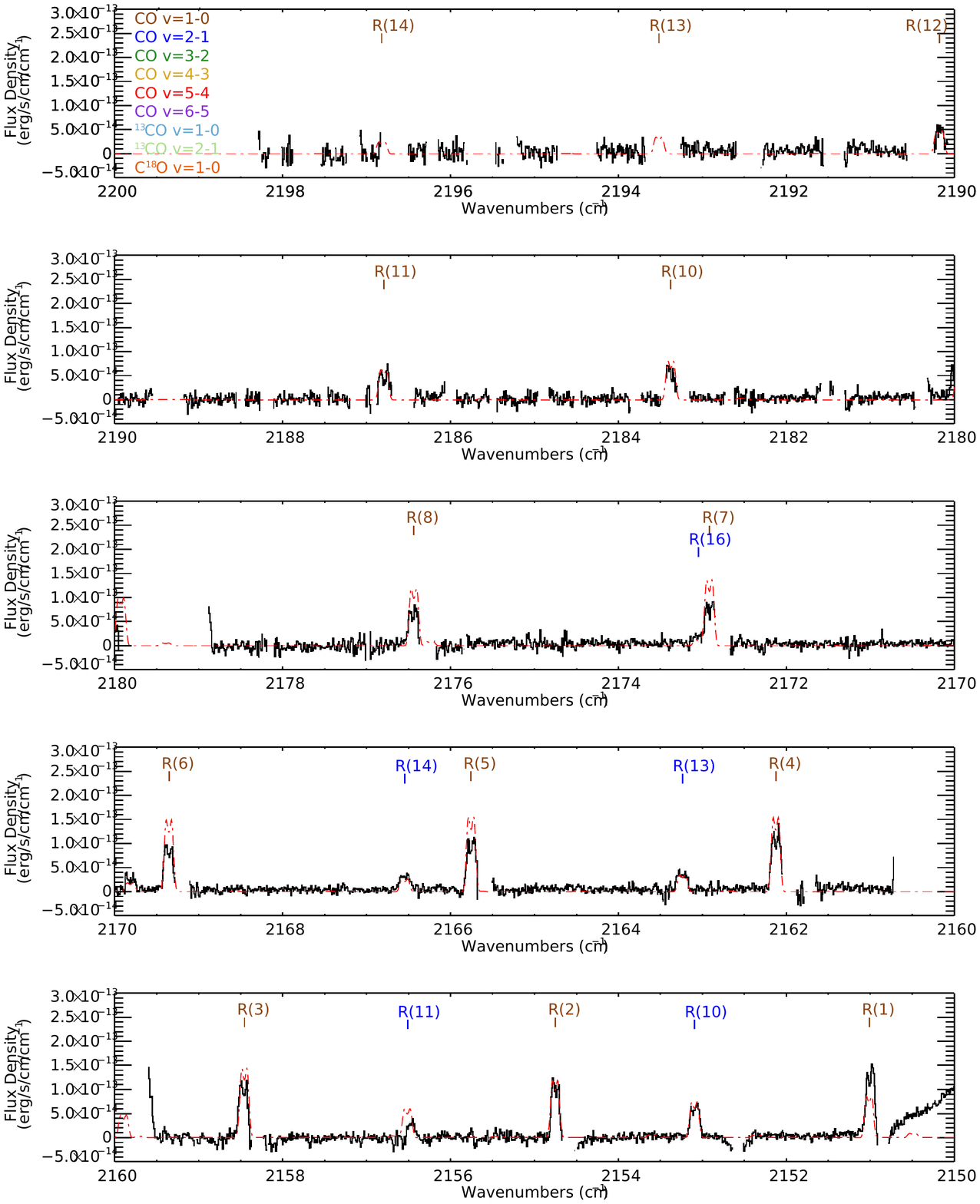}
\caption{\label{fig:part_spec_a} Partial spectrum for HD~141569 with model fit from 2200 to 2150 cm$^{-1}$. Black lines are data and dotted red lines are the final model. Regions without data are either segments where more than 40\% of flux was lost to absorption features or are gaps between orders. The locations of various ro-vibrational lines are marked. They are, from top to bottom, $^{12}$CO bands v=1-0 to 6-5, then $^{13}$CO bands v=1-0 and 2-1, and finally C$^{18}$O band v=1-0. The color of each band is indicated in the first panel.}
\end{figure}

\begin{figure}[htb!]
\includegraphics[trim=0 25 0 25, width=1.0\textwidth]{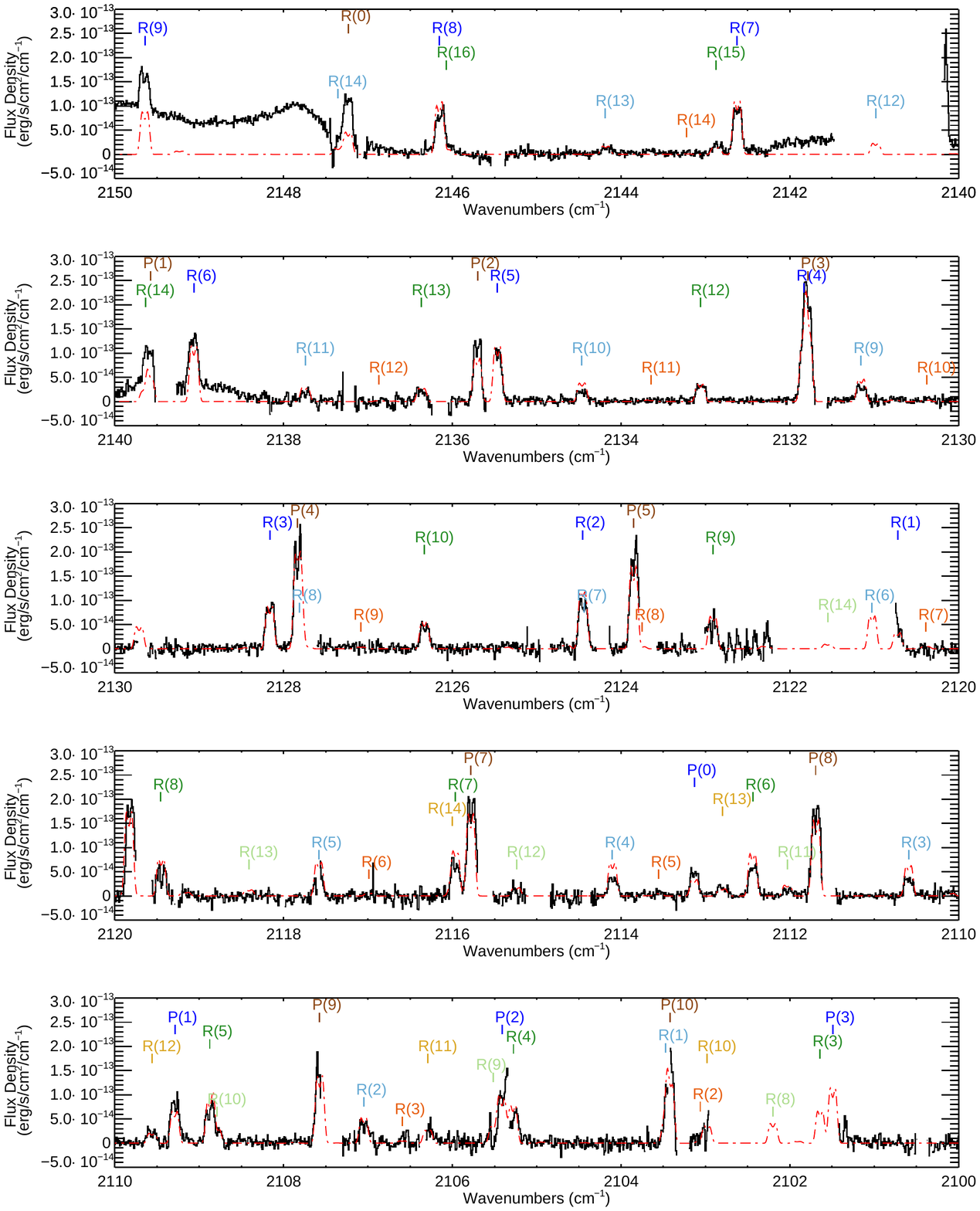}
\caption{\label{fig:part_spec_b} Similar to Figure \ref{fig:part_spec_a} for the wavenumber range of 2150 to 2100 cm$^{-1}$.}
\end{figure}

\begin{figure}[htb!]
\includegraphics[trim=0 25 0 25, width=1.0\textwidth]{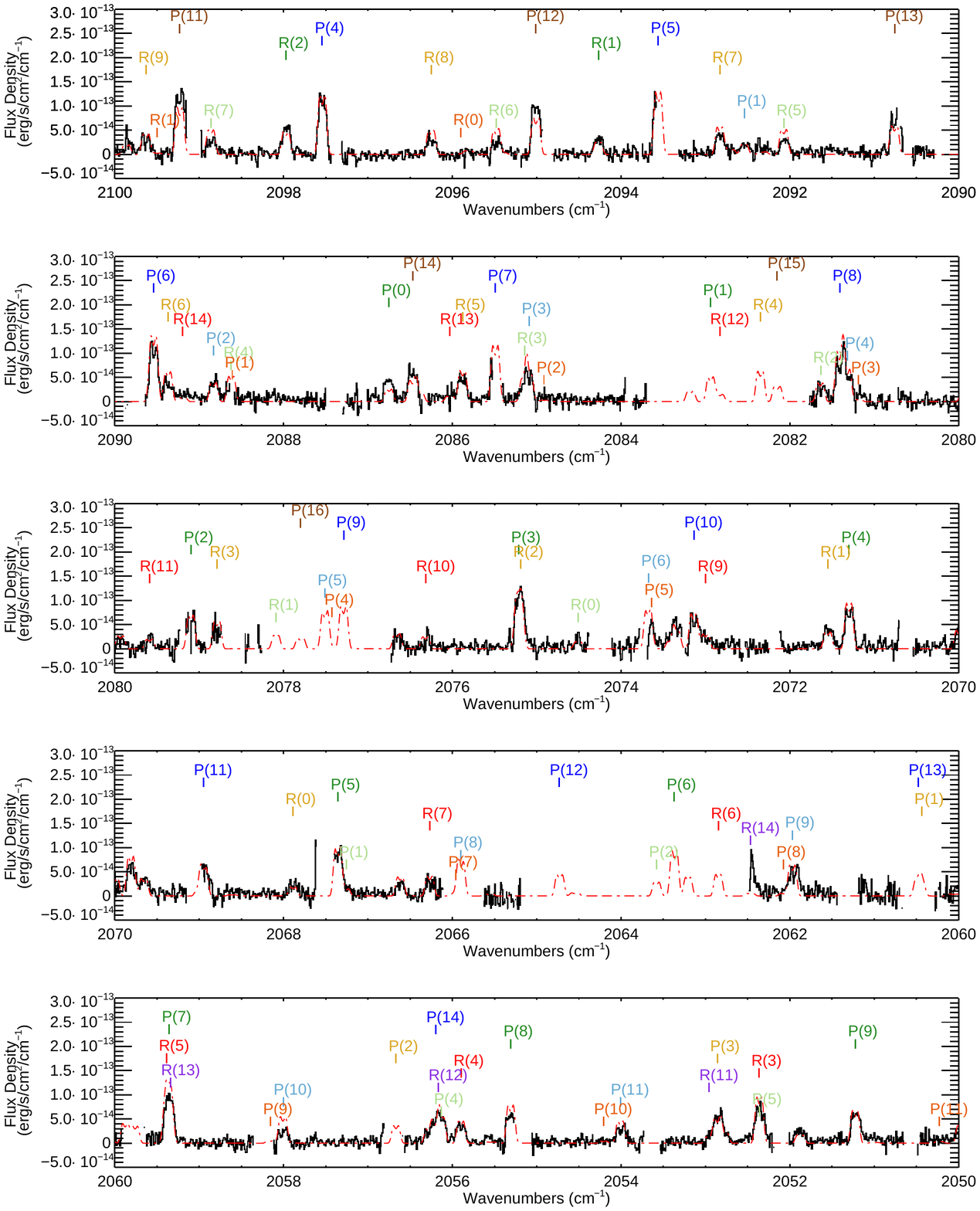}
\caption{\label{fig:part_spec_c} Similar to Figure \ref{fig:part_spec_a} for the wavenumber range of 2100 to 2050 cm$^{-1}$.}
\end{figure}

\begin{figure}[htb!]
\includegraphics[trim=0 25 0 25, width=1.0\textwidth]{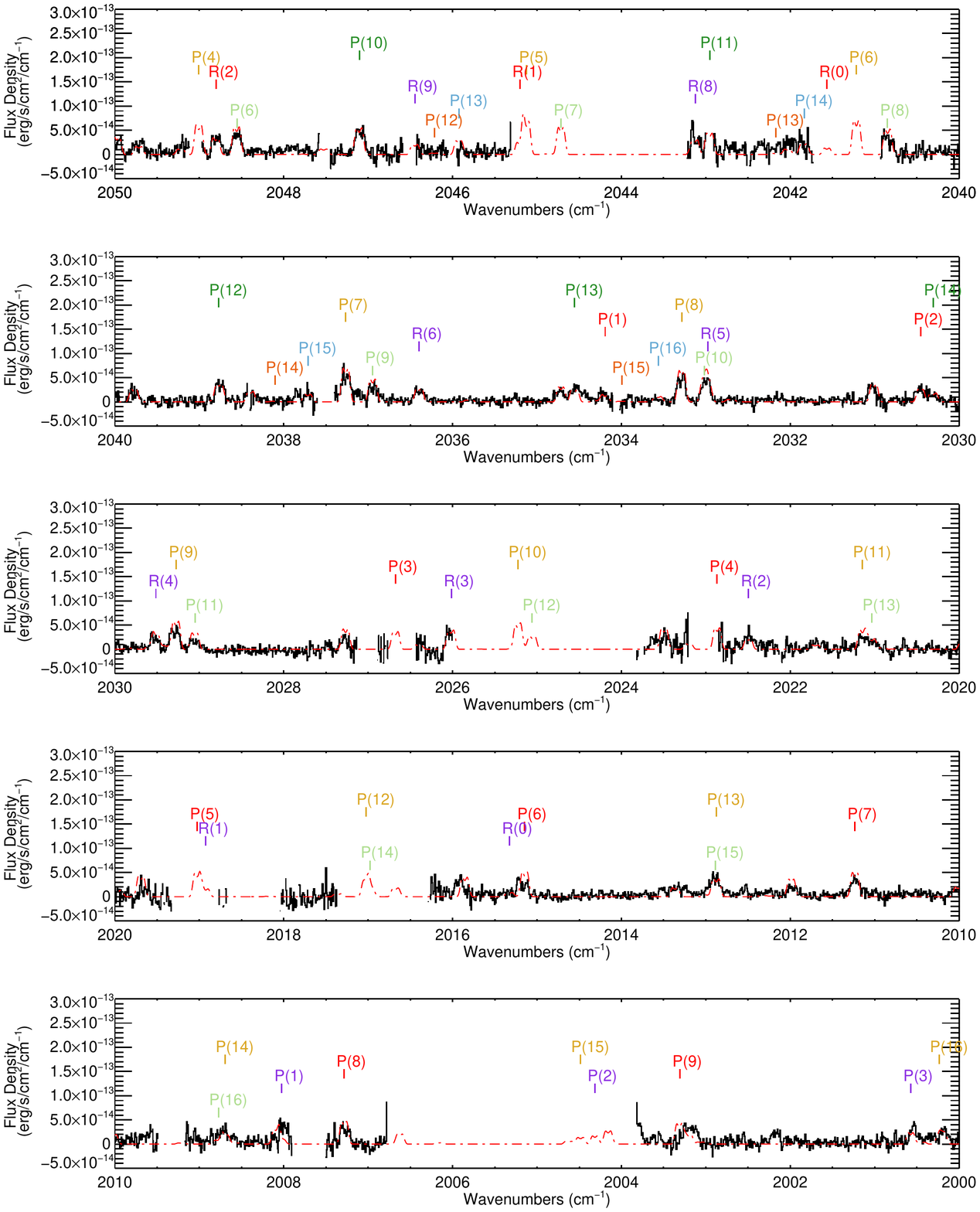}
\caption{\label{fig:part_spec_d} Similar to Figure \ref{fig:part_spec_a} for the wavenumber range of 2050 to 2000 cm$^{-1}$.}
\end{figure}

\begin{figure}[htb!]
\includegraphics[trim=0 25 0 25, width=1.0\textwidth]{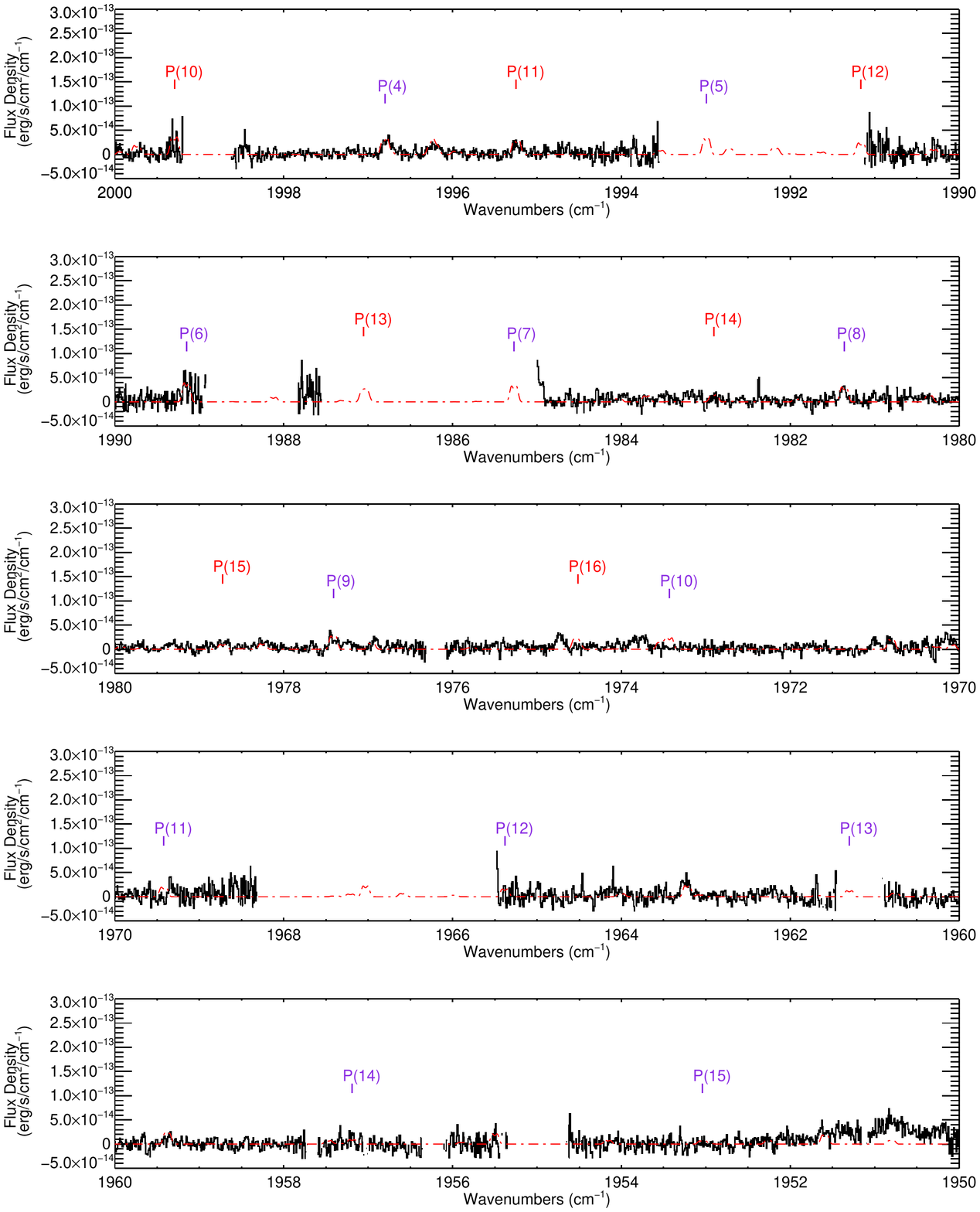}
\caption{\label{fig:part_spec_e} Similar to Figure \ref{fig:part_spec_a} for the wavenumber range of 2000 to 1950 cm$^{-1}$.}
\end{figure}

Once all three sets of data were reduced as described, the spectra and centroids were averaged together to create the final versions (Figs. \ref{fig:part_spec_a}, \ref{fig:part_spec_b}, \ref{fig:part_spec_c}, \ref{fig:part_spec_d}, \ref{fig:part_spec_e}, \ref{fig:part_sa_a}, \ref{fig:part_sa_b}, \ref{fig:part_sa_c}, \ref{fig:part_sa_d}, and \ref{fig:part_sa_e}).

\begin{figure}[htb!]
\includegraphics[trim=0 120 0 120,width=1.0\textwidth]{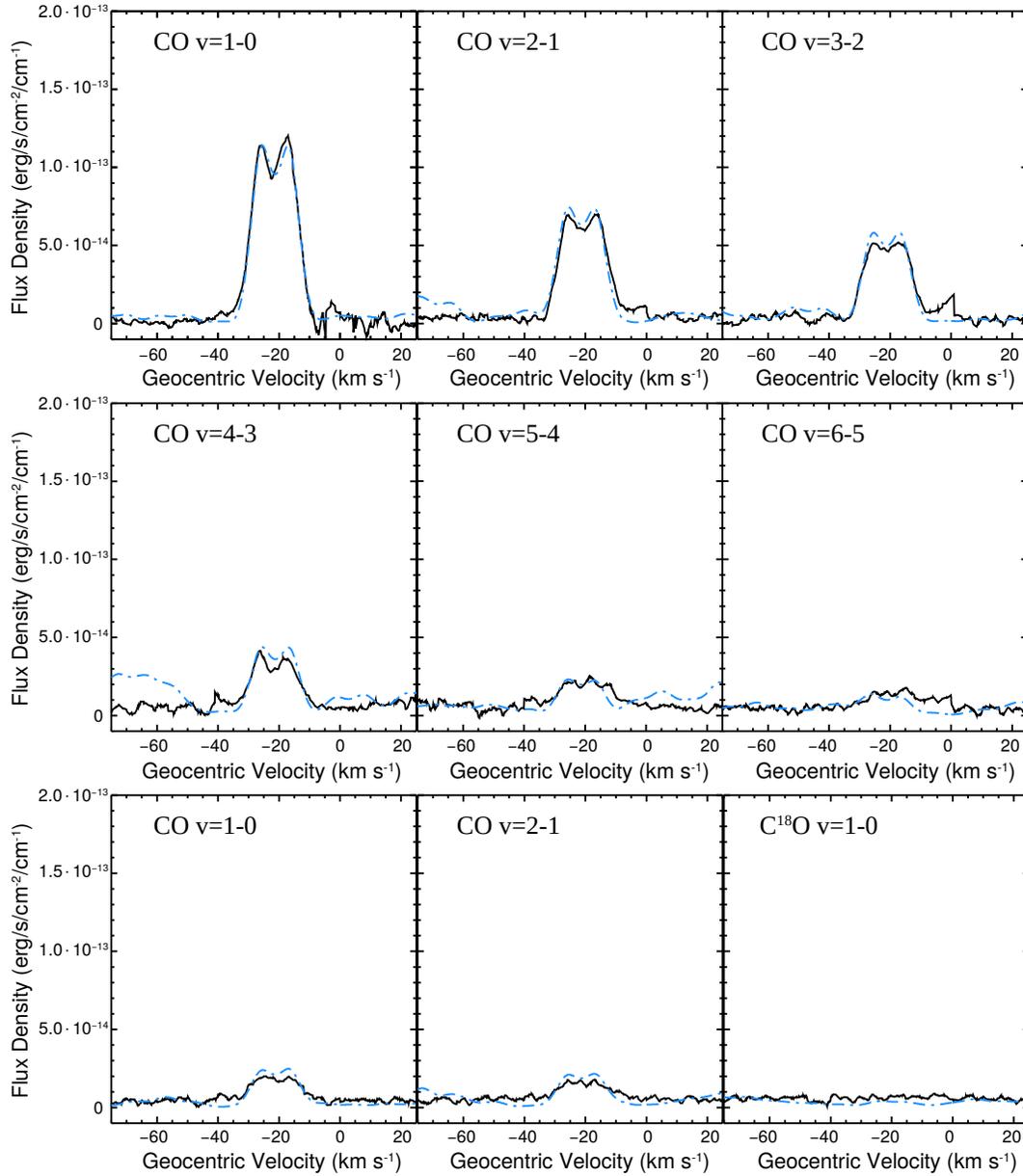}
\caption{\label{fig:stack_spec} Gallery of average spectroscopic line profiles for both the data and fluorescence model. The top row, from left to right, contains $^{12}$CO v=1-0, 2-1, and 3-2 vibrational bands. The middle row contains $^{12}$CO v=4-3, 5-4, and 6-5 bands. The bottom row contains $^{13}$CO v=1-0 and 2-1 bands, and the C$^{18}$O v=1-0 band. In all cases the data profile is in black and the model profile in blue. The apparent excess in the $^{12}$CO v=4-3 plot is the result of a large feature unrelated to the vibrational band which was not completely averaged out during line stacking.}
\end{figure}

Spectral and spectroastrometric CO line profiles were created by stacking individual lines within each vibrational band in order to improve the signal-to-noise ratio. 
Figures \ref{fig:stack_spec} and \ref{fig:stack_sa} show the spectroscopic and spectroastrometric average profiles along with a model to be elaborated on in section \ref{modeling}.

\section{Results}
\begin{figure}[htb!]
\includegraphics[trim=0 25 0 25, width=1.0\textwidth]{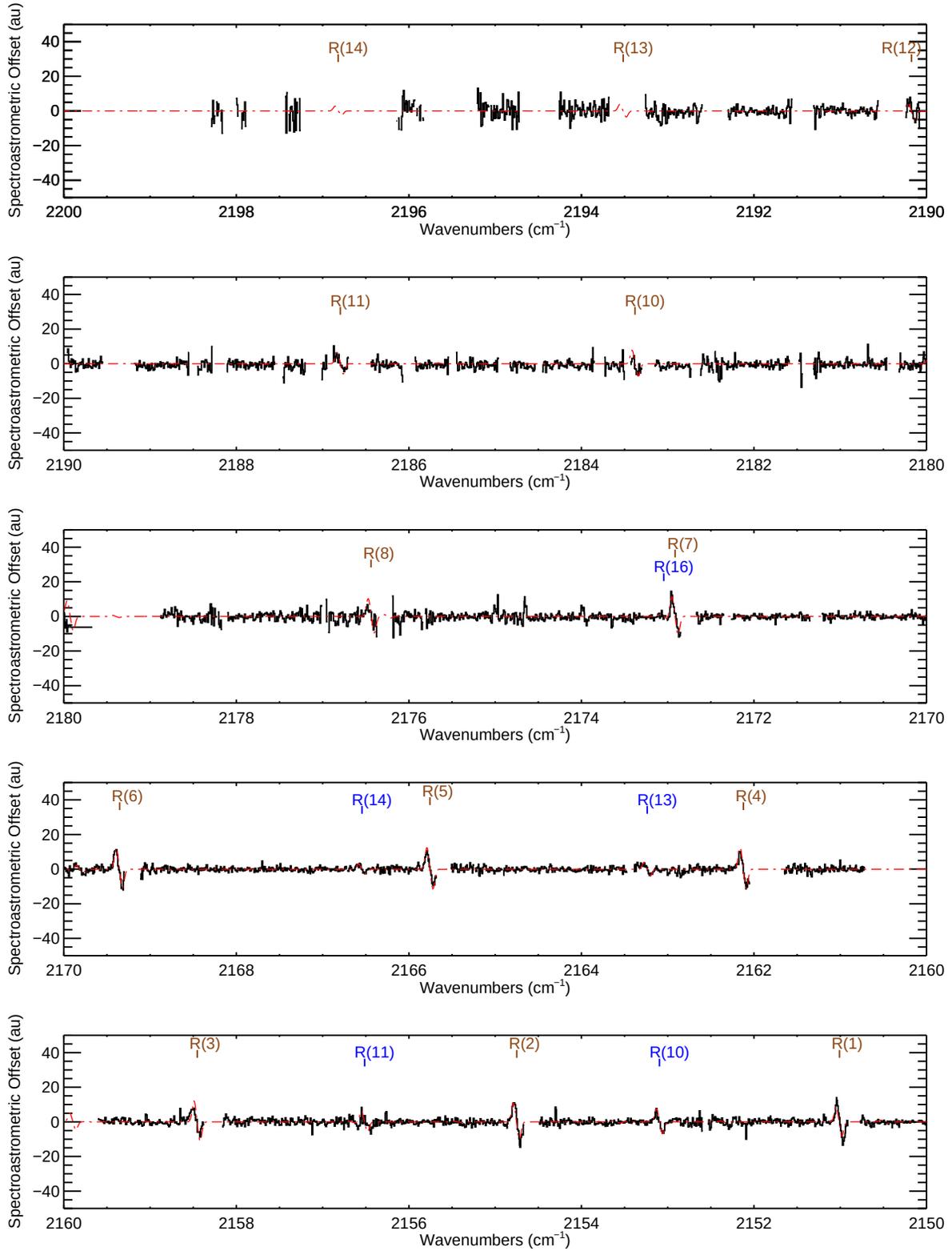}
\caption{\label{fig:part_sa_a} Partial spectroastrometric data for HD~141569 with model fit fit from 2200 to 2150 cm$^{-1}$. Black lines are data and dotted red lines are the final model. Regions without data are segments where more than 40\% of flux was lost to absorption features. The locations of various ro-vibrational lines are marked. They are, from top to bottom, $^{12}$CO bands v=1-0 to 6-5, then $^{13}$CO bands v=1-0 and 2-1, and finally C$^{18}$O band v=1-0. The same color scheme used in figure 1 is used here.}
\end{figure}

\begin{figure}[htb!]
\includegraphics[trim=0 25 0 25, width=1.0\textwidth]{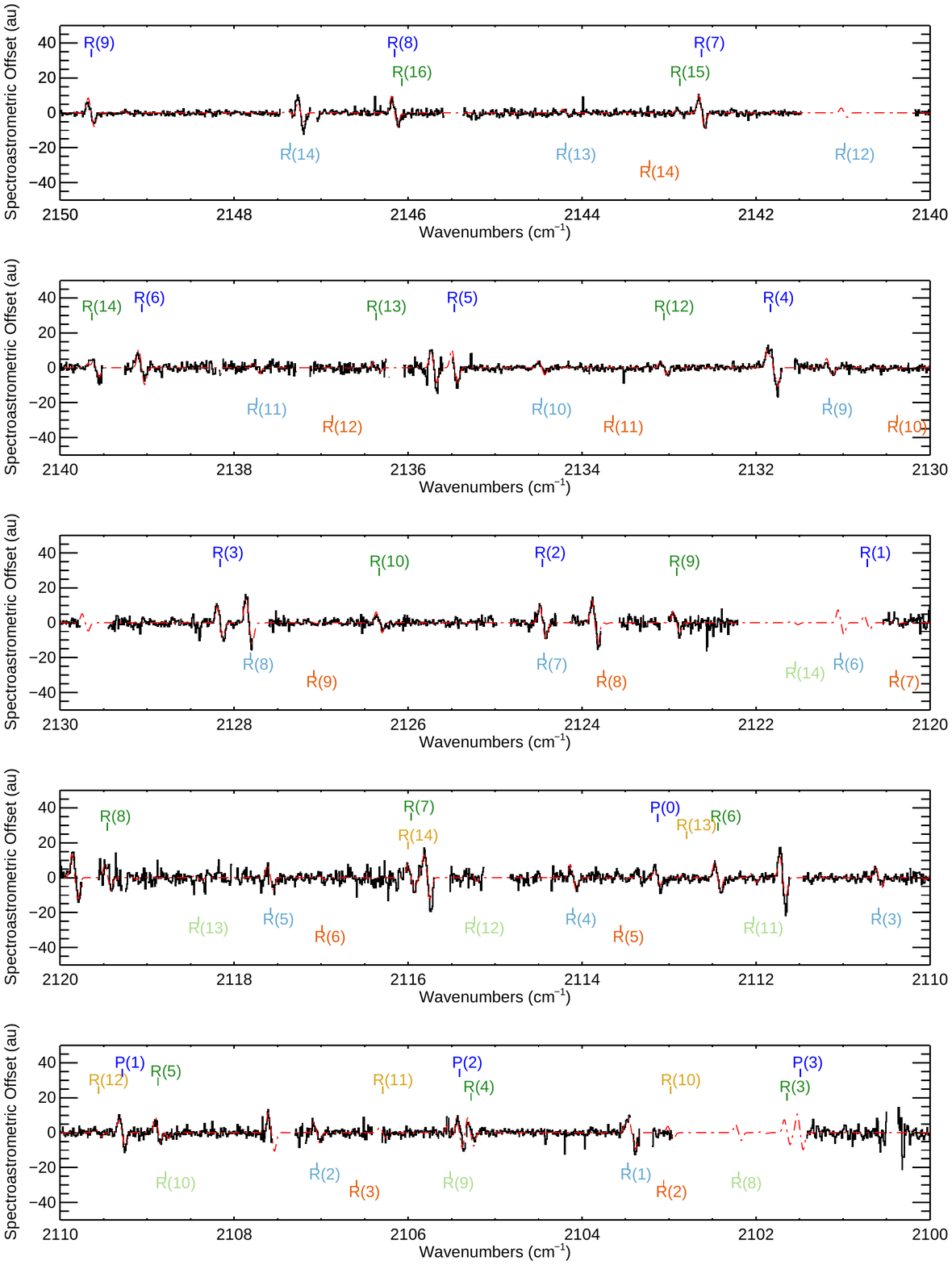}
\caption{\label{fig:part_sa_b} Similar to Figure \ref{fig:part_sa_a} for the wavenumber range of 2150 to 2100 cm$^{-1}$.}
\end{figure}

\begin{figure}[htb!]
\includegraphics[trim=0 25 0 25, width=1.0\textwidth]{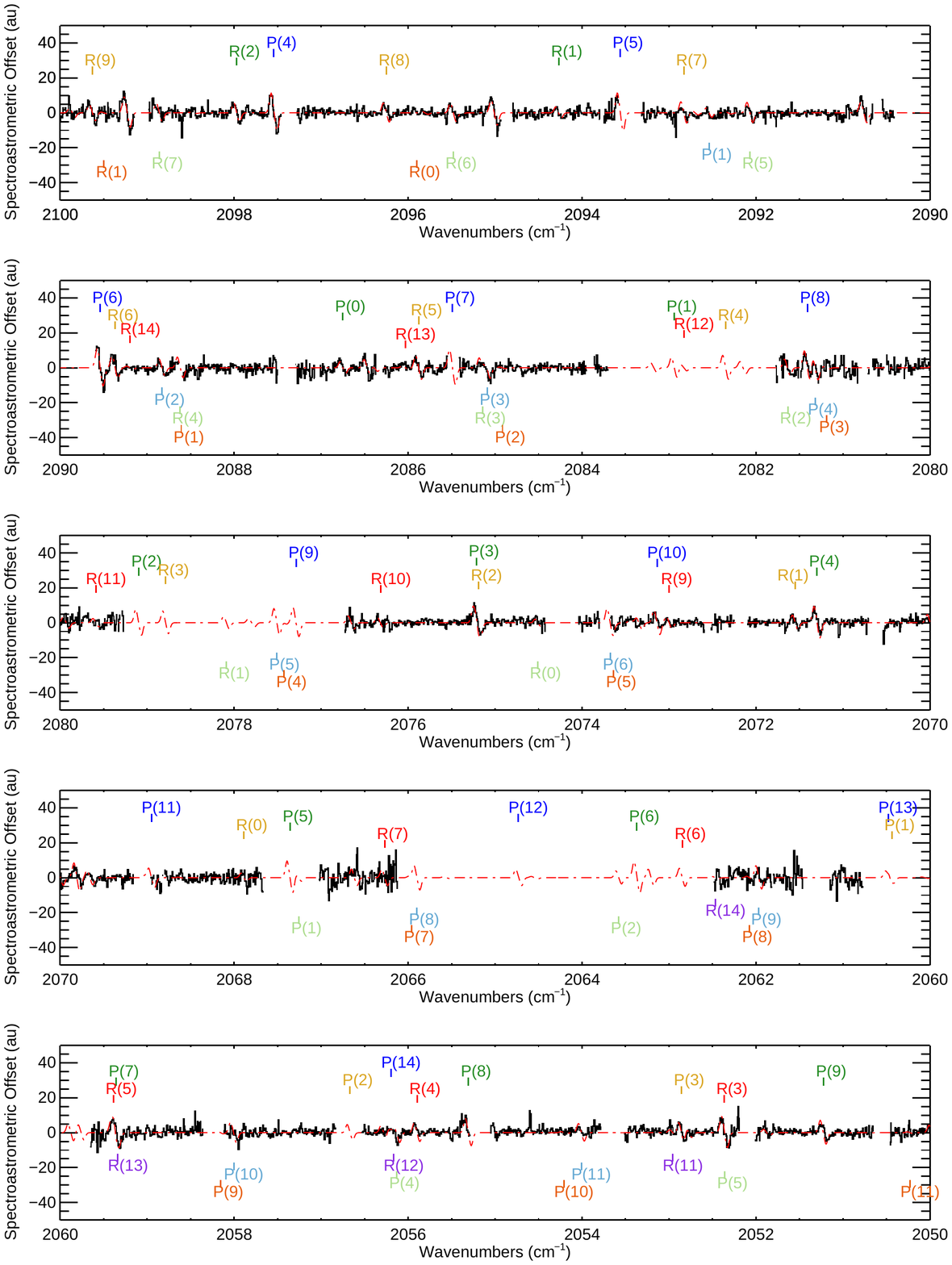}
\caption{\label{fig:part_sa_c} Similar to Figure \ref{fig:part_sa_a} for the wavenumber range of 2100 to 2050 cm$^{-1}$.}
\end{figure}

\begin{figure}[htb!]
\includegraphics[trim=0 25 0 25, width=1.0\textwidth]{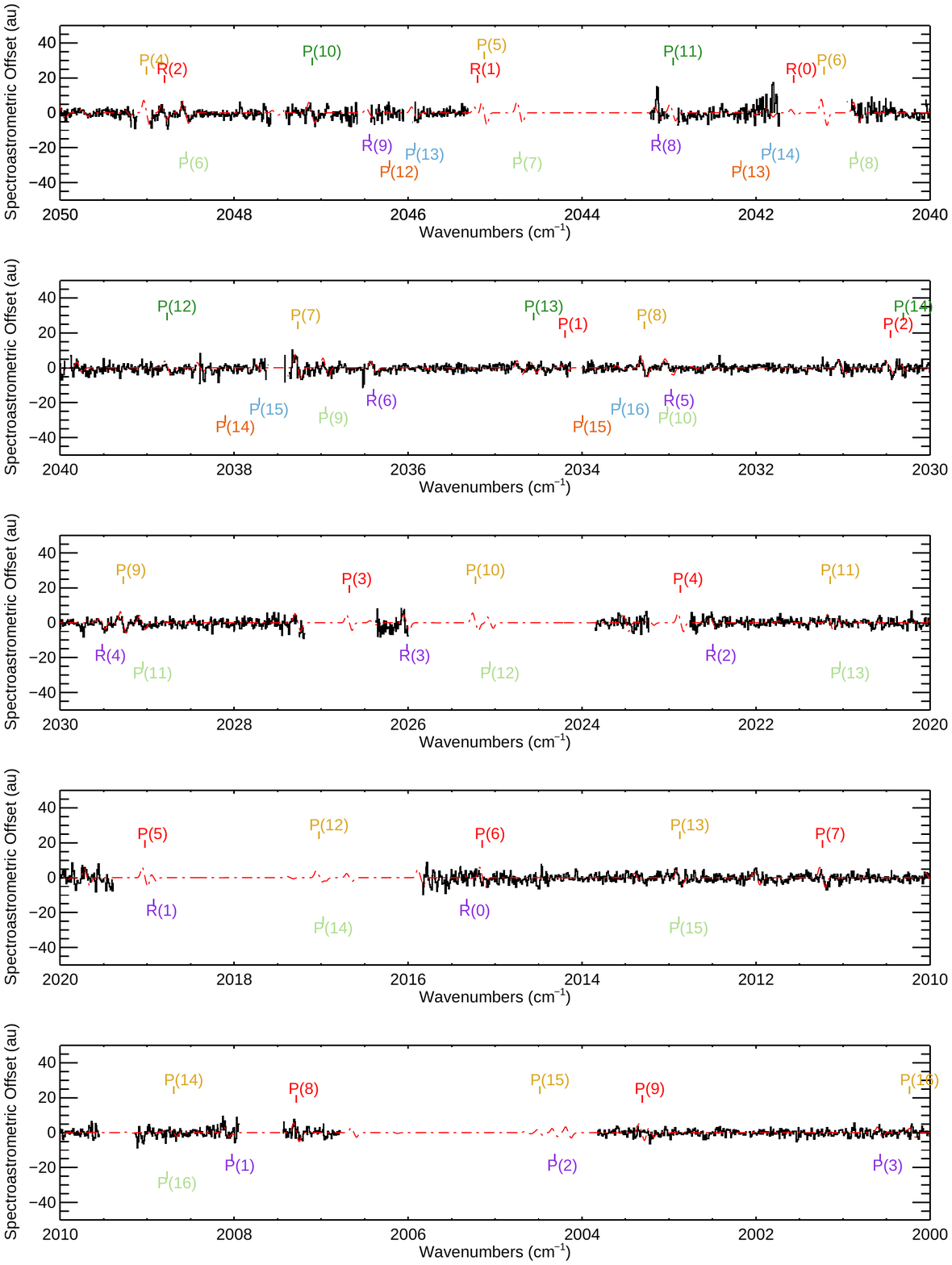}
\caption{\label{fig:part_sa_d} Similar to Figure \ref{fig:part_sa_a} for the wavenumber range of 2050 to 2000 cm$^{-1}$.}
\end{figure}

\begin{figure}[htb!]
\includegraphics[trim=0 25 0 25, width=1.0\textwidth]{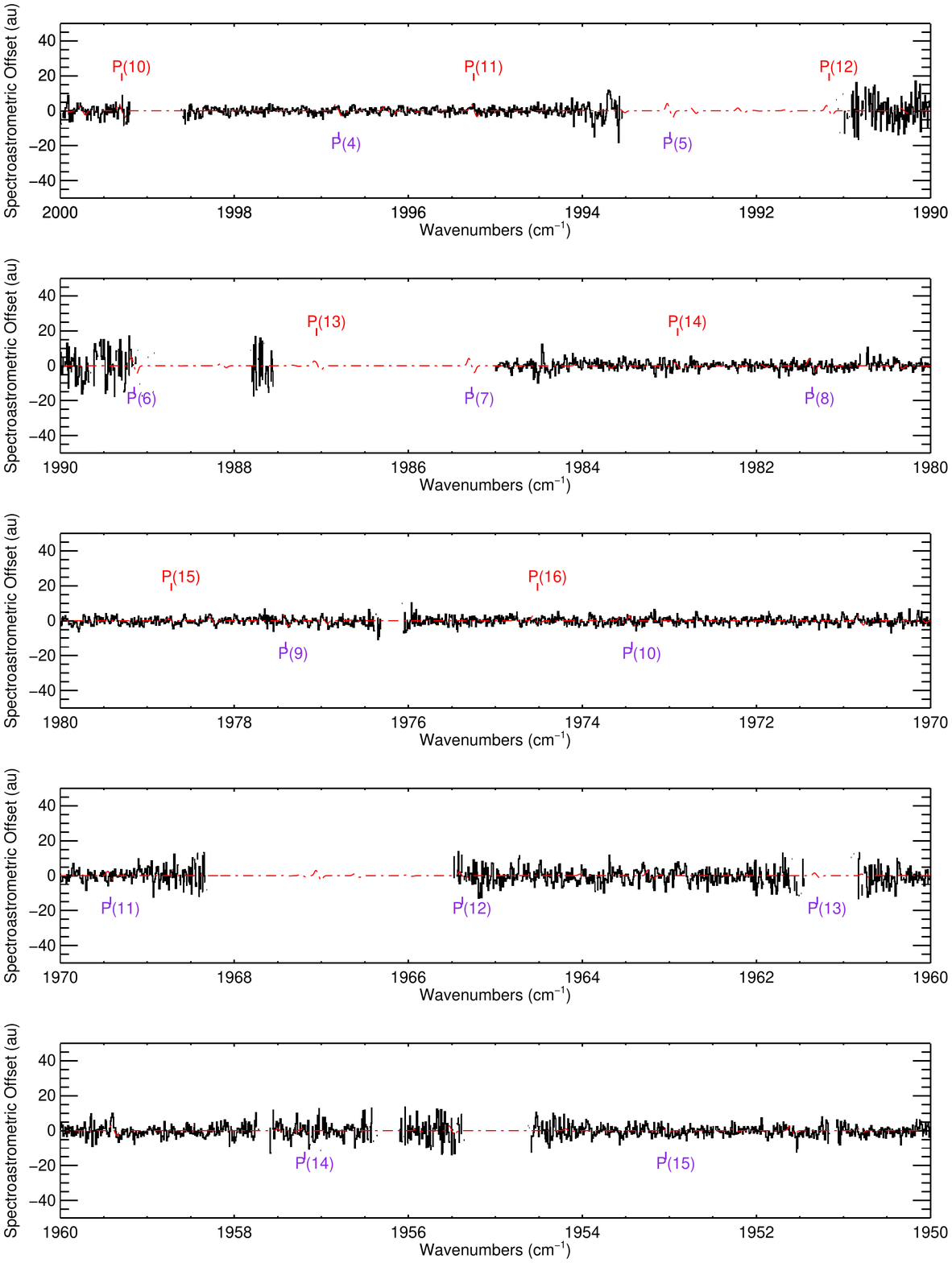}
\caption{\label{fig:part_sa_e} Similar to Figure \ref{fig:part_sa_a} for the wavenumber range of 2000 to 1950 cm$^{-1}$.}
\end{figure}

The wavenumber range from 1950 to 2200 cm$^{-1}$ (approximately 4.5 to 5.1 $\mu$m) was examined, which provides ample access to numerous CO ro-vibrational lines. This includes coverage of isotopolgues such as $^{13}$CO and C$^{18}$O, though the latter was not observed here. The non-detection of C$^{18}$O has previously been noted in ALMA observations by \cite{miley18}, who suggested selective photodissociation for the isotopologue as a possible cause. Figures \ref{fig:part_spec_a}, \ref{fig:part_spec_b}, \ref{fig:part_spec_c}, \ref{fig:part_spec_d}, \ref{fig:part_spec_e}, \ref{fig:part_sa_a}, \ref{fig:part_sa_b} \ref{fig:part_sa_c}, \ref{fig:part_sa_d}, and \ref{fig:part_sa_e} show the spectrum and centroid for HD~141569. The stronger $^{12}$CO v=1-0 and v=2-1 emission features show a clear double-peaked profile. $^{12}$CO lines from up to v=5-4 vibrational transitions are readily visible in the spectrum. A few v=6-5 and v=7-6 lines were found as well, though often with confounding aspects such as high local noise and possible blending with weak lines. Additional lines are also evident from $^{13}$CO v=1-0 and v=2-1 transitions. Any possible lines from additional CO isotopologues or vibrational bands are either too weak to identify reliably or are blended with much stronger features. When measuring the equivalent widths, lines were ignored if they showed signs of being strongly blended, or having lost significant portions to telluric absorption. The measured equivalent widths are shown in Table \ref{tab:eqw_long}. 

\begin{figure}[htb!]
\includegraphics[trim=0 120 0 120,width=1.0\textwidth]{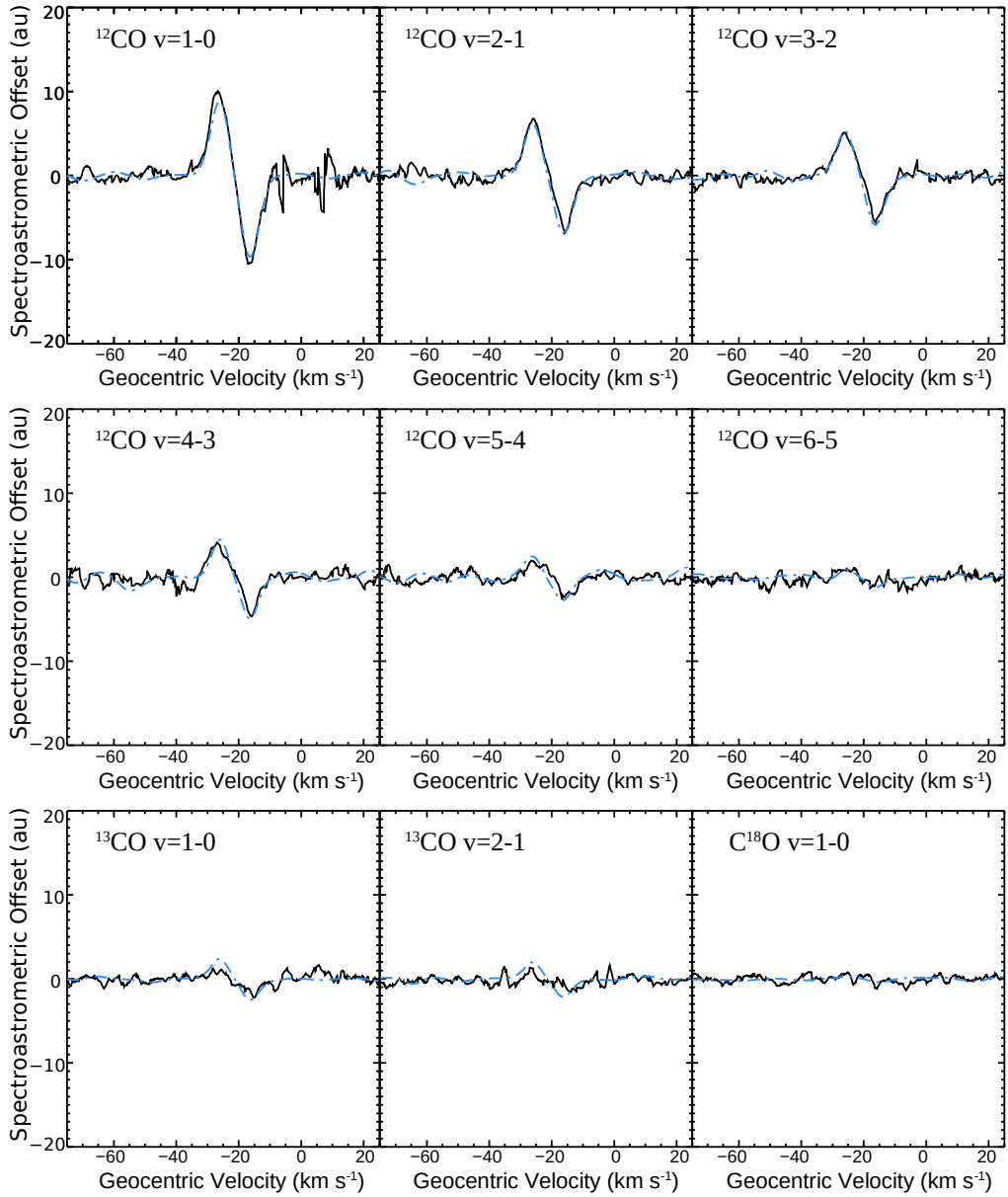}
\caption{\label{fig:stack_sa} Similar to Fig. \ref{fig:stack_spec} with the average spectroastrometric profile for the various vibrational bands. The top row, from left to right, contains $^{12}$CO v=1-0, 2-1, and 3-2 vibrational bands. The middle row contains $^{12}$CO v=4-3, 5-4, and 6-5 bands. The bottom row contains $^{13}$CO v=1-0 and 2-1 bands, and the C$^{18}$O v=1-0 band. In all cases the data profile is in black and the model profile in blue.}
\end{figure}

The line profiles obtained by line stacking provide some immediate insight. These can be seen in Figure \ref{fig:stack_spec} for the spetroscopic profiles, and Figure \ref{fig:stack_sa} for the spectroastrometric profiles. The $^{12}$CO v=1-0 profile is symmetric. Depending on the adopted stellar mass, here the assumed range varies from 2 to 2.4 M$_\odot$ in line with previous results such as those from \cite{merin04} and \cite{white16}, respectively, the wings at HWZI ($\sim$10 km s$^{-1}$) imply an inner edge to the emitting region of approximately 12-15 au, and the two peaks an outer edge of 50-90 au for Keplerian orbits.

The spectroastrometric signal, from by-eye inspection of individual lines (Figures \ref{fig:part_sa_a}, \ref{fig:part_sa_b}, \ref{fig:part_sa_c}, \ref{fig:part_sa_d}, and \ref{fig:part_sa_e}), seems indicative of primarily or solely axisymmetric 
Keplerian rotation of the disk. This is reinforced, again through by-eye inspection, in the line stacking (Figure \ref{fig:stack_sa}) which also produces the characteristic "s"-shaped curve of differential rotation by an axisymmetric 
Keplerian disk.

Assuming that the gas responsible for the observed emission is optically thin, the population of each level is related to the line flux by
\begin{equation}
    F_{ij} = \frac{h c \tilde{\nu}_{ij} A_{ij} N_j }{4 \pi d^2}
    \label{eq:line_flux}
\end{equation}
where $h$ is the Planck constant, $c$ the speed of light, $\tilde{\nu}_{ij}$ the wavenumber of the transition between states $i$ and $j$, $A_{ij}$ the Einstein coefficient, $N_j$ the population of state $j$, and $d$ is the distance to the object. The relative population of the rotational levels is
\begin{equation}
    N_j = \frac{g_j N e^{ -\frac{E'}{k_B T} } }{Q}
    \label{eq:rel_pop}
\end{equation}
where $g_j$ is the degeneracy of state $j$, $N$ is the source population, $E'$ is the energy of the upper level, $k_B$ is the Boltzmann constant, $T$ is the temperature, and $Q$ is the partition function. This can be rearranged into a linear form such that
\begin{equation}
    \ln\left(\frac{N_j}{g_j}\right) = -\frac{E^\prime j}{k_{B} T} + \ln\left(\frac{N}{Q}\right).
    \label{eq:lin_prop}
\end{equation}
The vibrational lines follow a similar form.

\begin{figure}[htb!]
\includegraphics[trim=50 140 50 200,width=1.\textwidth]{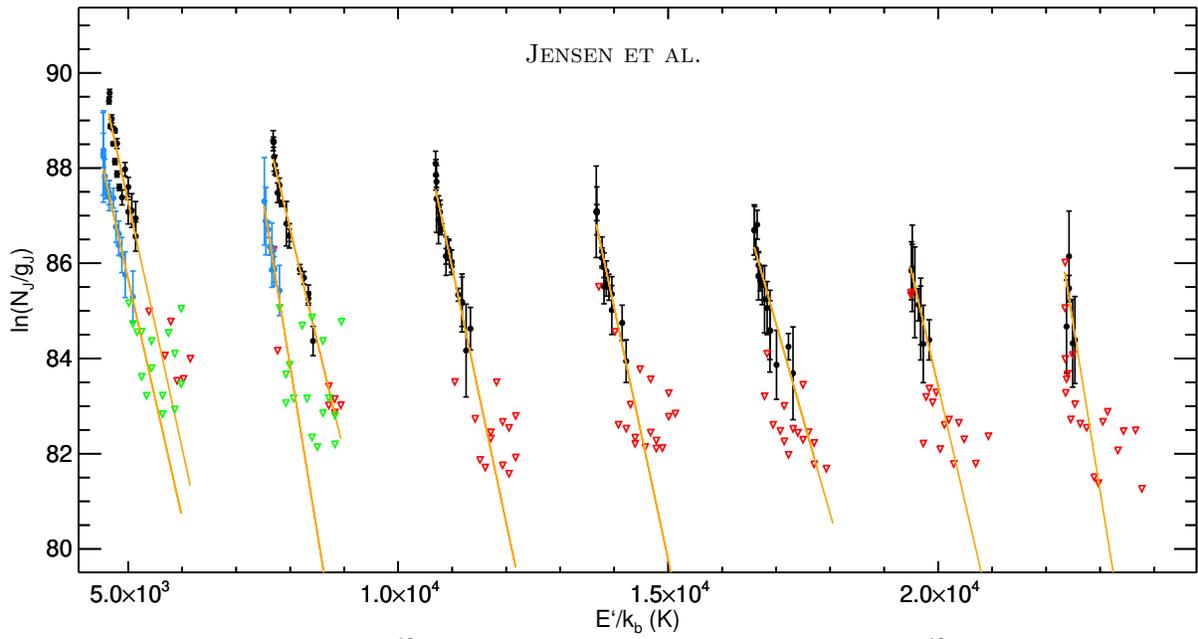}
\caption{\label{fig:rotex_all} Excitation diagram for CO lines.  $^{12}$CO measurements are shown in filled black circles. $^{13}$CO measurements in blue. Upper limits are given by downward-pointing triangles, red for $^{12}$CO and green for $^{12}$CO. Vibrational band increases from left to right starting with v=1-0. The fits, orange, are those derived from the results of the method of \cite{kelly07}.}
\end{figure}

For the linear fitting, we adopt the likelihood-based method of \cite{kelly07} through the IDL code \texttt{LINMIX\_ERR} due to its treatment of both measurement errors and upper limits. The results for majority of the vibrational bands observed for $^{12}$CO are consistent at approximately 200~K. However, the derived rotational temperature from the upper-limit-dominated v=7-6 band is highly uncertain. The results from all the fits are summarized in Table \ref{tab:rottemp}. The rotational fits are likewise shown in Figure \ref{fig:rotex_all}.

The vibrational temperature, using simple linear regression with the results of the method of \cite{kelly07}, is 4540$\pm$440~K. The high vibrational temperature compared to its rotational counterpart is indicative of UV fluorescence and consistent with the value of 5600$\pm$800~K found by \cite{brittain07}. The new result incorporates more lines at a higher resolution, which allows for better line separation and thus a smaller uncertainty in the vibrational temperature.

\section{Modeling} \label{modeling}
In examining the spectrum of HD~141569, two separates models were used. The first is a 1+1D slab fluorescence model, of the kind used by \cite{brittain09}, that generates CO emission features from a disk of given parameters. The disk itself is assumed to be Keplerian and axisymmetric in nature as well as vertically isothermal. The radial rotational temperature is described by a power law. A constant turbulent velocity was included for the material in the disk. The UV luminosity parameter, which determines the amount of ultraviolet emission from the star illuminating the disk, was also included to account for UV fluorescence in the disk. The relative fluxes of the isotopologues considered ($^{12}$CO, $^{13}$CO, and C$^{18}$O) are determined by scaling factors rather than explicitly taking into account effects such as selective dissociation or dust opacity (see, e.g., \citealt{brittain07,brittain09} for more details).

Parameters for the fluorescence model were iteratively determined by individual single-model runs as well as by $\chi ^2$ minimization through running multiple models. Single-model instances were primarily used for rough determination of parameter space and incremental adjustments between automated runs. The $\chi ^2$ minimization runs generated random combinations within the parameter space, compared the model results with the data, and returned the $\chi ^2$ values and the best fit model spectrum and centroid. These results were used to inform parameter space adjustments and the process repeated.

Each $\chi ^2$ minimization run allowed for variation of multiple parameters. Free parameters included inner and outer radii of emitting region, fiducial rotational temperature, the slope of the radial rotational temperature dependence $\alpha_{rot}$, fiducial hydrogen density, slope of the radial hydrogen density dependence $\alpha_H$, UV luminosity, turbulence velocity, and number of layers, where fiducial parameters were taken as the relevant parameter value at 1 au and the $\alpha$-values are the power law indices for the radial distribution of the respective quantities. The UV luminosity is based on extinction corrected data from the International Ultraviolet Explorer (IUE) (\cite{brittain07} and references therein). The original data assumed a distance of 99 pc. In this paper, we adopt the distance inferred from {\it Gaia} measurements 110.6 pc \citep{gaiamission,gaiadr2} and the flux increased by twenty percent. 

The number of layers was set to 7000, corresponding to a total column density of $1.9\times10^{17}$ cm$^{-2}$, at which point the gas grows optically thick to the UV radiation, for the material excited by UV emission along the star-disk line of sight. Accounting for the inclination of the disk, the column density along the observed line of sight is $5.3\times10^{16}$ cm$^{-2}$. Additional parameters were occasionally included, such as stellar mass, inclination, inner wall area, and disk PA relative to slit. These latter parameters were typically informed by various literature values or to assess their effect on the fit. The stellar mass was ultimately set at 2.39 M$_{\odot}$, consistent with \cite{white16}, and the disk PA was set as aligned with the slit. The final inclination was fixed to 51$^{\circ}$, in line with \cite{weinberger99}. In the case of $\chi ^2$ minimization runs, the parameters were generated randomly within ranges based on values found in the literature, results of individual model iteration, and results of other minimization runs.

The final parameters include an emission region with 13$\pm$2 au and 60$\pm$5 au for the inner-outer radii, respectively, a fiducial rotational temperature of 495$\pm$5~K at 1 au, and a temperature $\alpha_{rot}$ value of 0.275$\pm$0.020. The full line list can be found in Table \ref{tab:eqw_long}. Average profiles for both the spectrum and spectroastrometric signal of the $^{12}$CO, $^{13}$CO, and C$^{18}$O lines can be seen in Figs. \ref{fig:stack_spec} and \ref{fig:stack_sa}.

\begin{figure}[htb!]
\includegraphics[trim=0 150 0 150, width=1.0\textwidth]{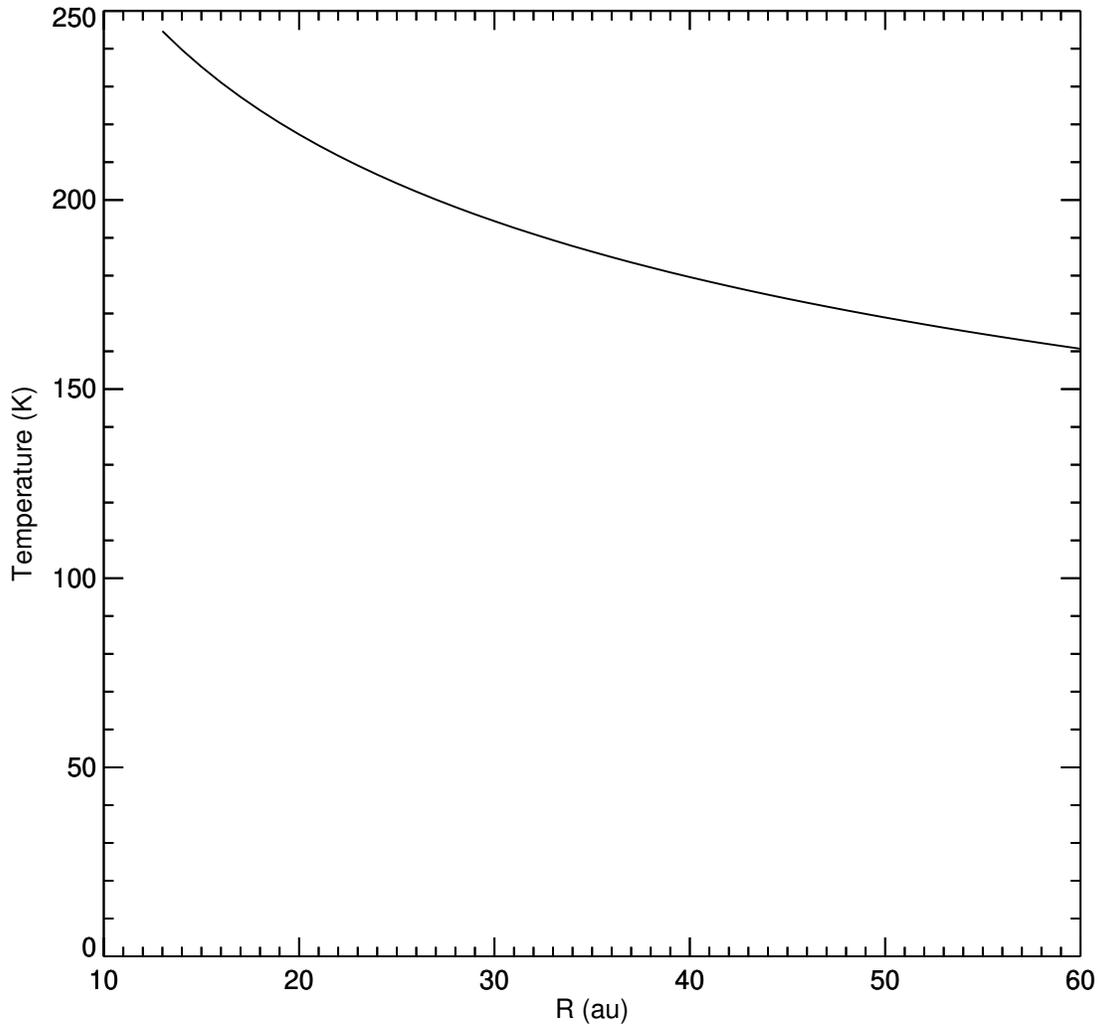}
\caption{\label{fig:mod_temp} Final model temperature as a function of radius. Calculated as a power law, with fiducial (1 au) temperature of 495~K and exponent of 0.275.}
\end{figure}

\begin{figure}[htb!]
\includegraphics[trim=0 150 0 150, width=1.0\textwidth]{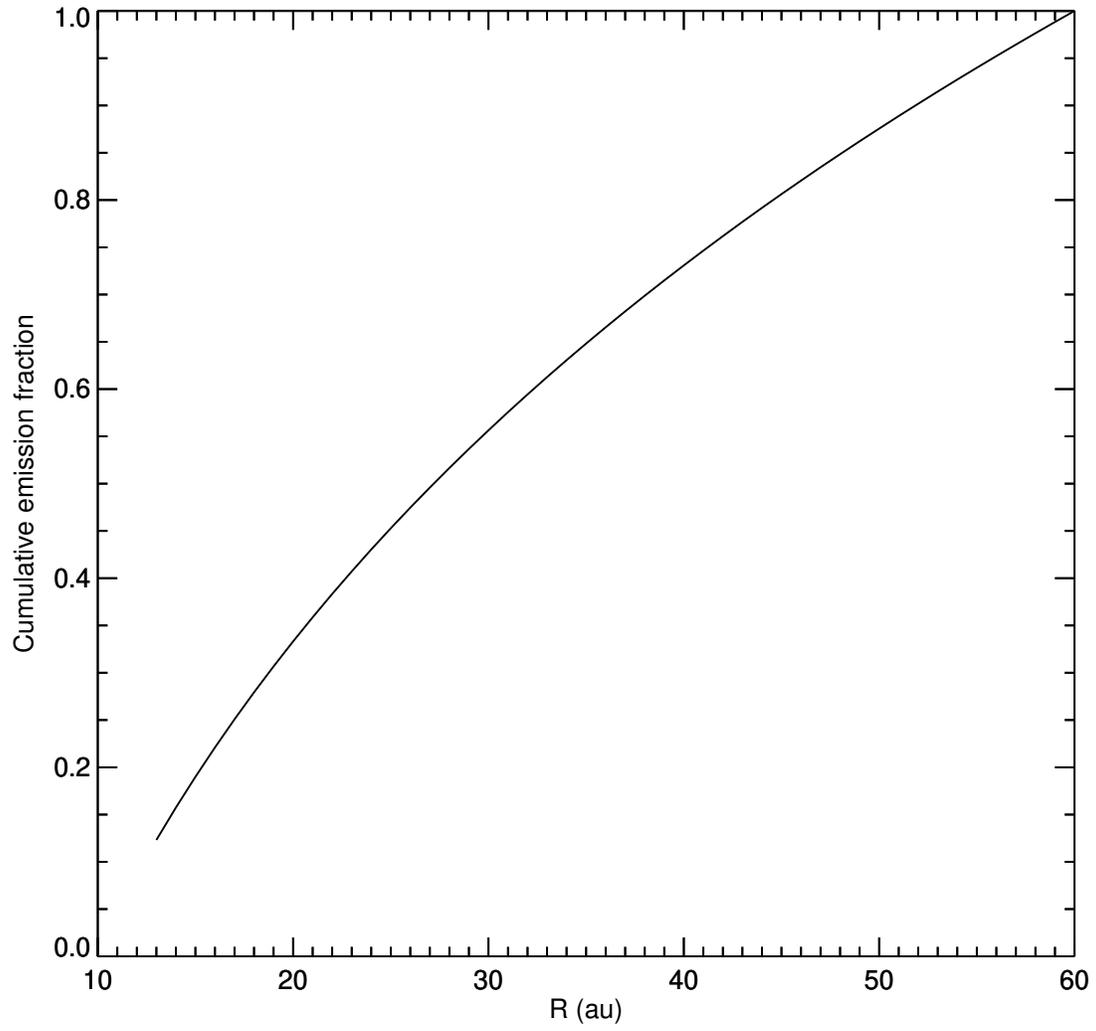}
\caption{\label{fig:cumulum} Final model cumulative luminosity as a function of radius.}
\end{figure}

\begin{figure}[htb!]
\includegraphics[trim=0 150 0 150, width=1.0\textwidth]{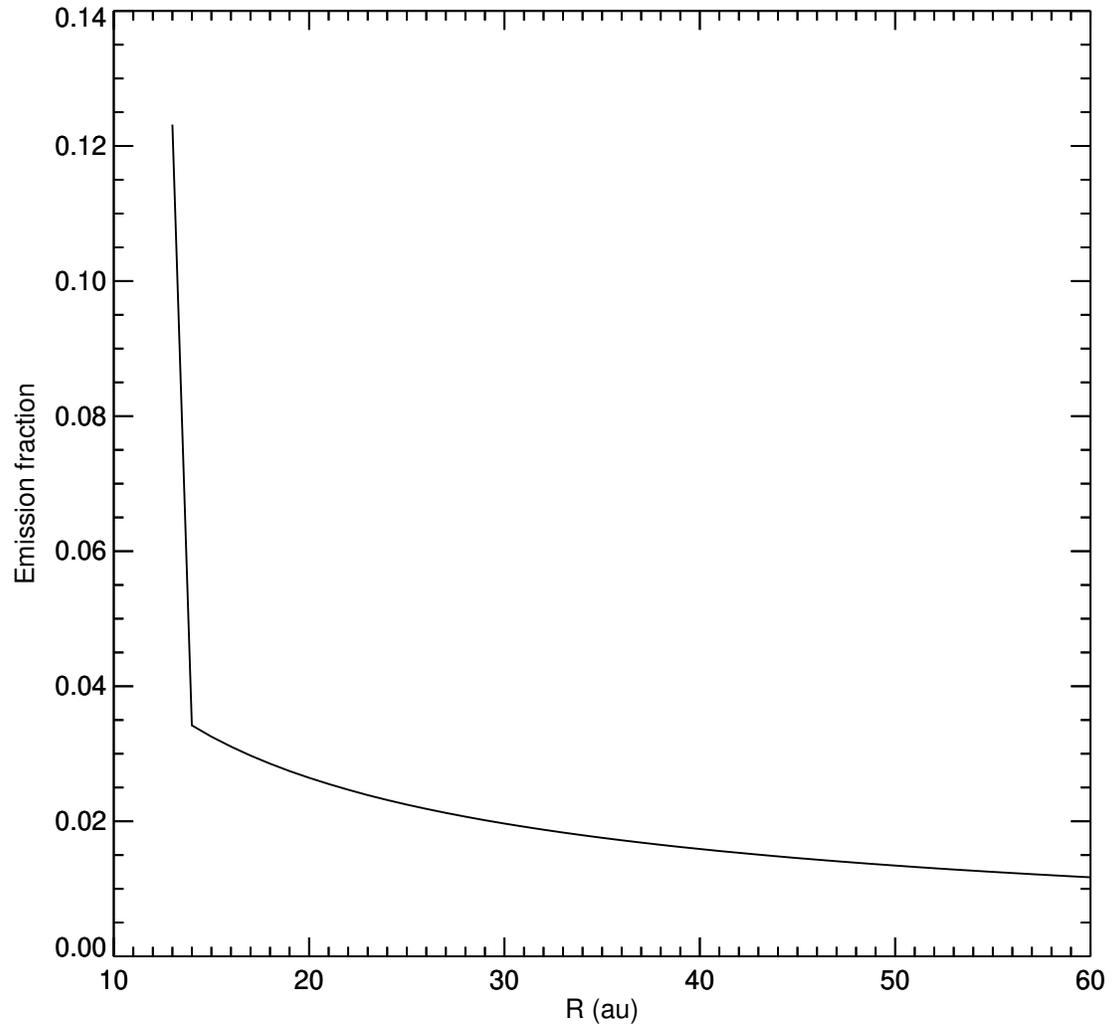}
\caption{\label{fig:lum_prof} Final model total emission fraction as a function of radius.}
\end{figure}

The fluorescence model's temperature profile as a function of radius dips below the average rotational temperature of $\sim$200 K, as can be seen in Fig. \ref{fig:mod_temp}. This likewise would bring the annulus-averaged temperature well below that established temperature. However, as can be seen in Figs. \ref{fig:cumulum} and \ref{fig:lum_prof} the majority of the emission is coming from the inner regions. When the temperature is weighted by the annuli luminosity, the average temperature is found to be 202~K.

The line profiles of warm gas in a disk may be asymmetric for various reasons, including emission from a circumplanetary disk or emission from an eccentric disk (e.g. \citealt{kley06,regaly2011,liskowsky12}). In order to further verify that there is no evidence of asymmetry in the line profile, a second model for eccentric disks was used. The eccentric disk model is much simpler than the fluorescence model previously described. The disk in this model has an eccentric inner rim, with the azimuthally averaged eccentricity decreasing with radius. The geometry of the disk (eccentricity profile, inner and outer radii), stellar mass, and radial intensity profile are the free parameters. Rather than generating an entire spectrum as in the fluorescence model only a single line profile in velocity space was produced per run. This profile was compared with the average profile from observations, with the final fit determined through $\chi ^2$ minimization.

\begin{figure}[htb!]
    \centering
    \includegraphics[trim=70 35 70 70 ,width=1.\textwidth]{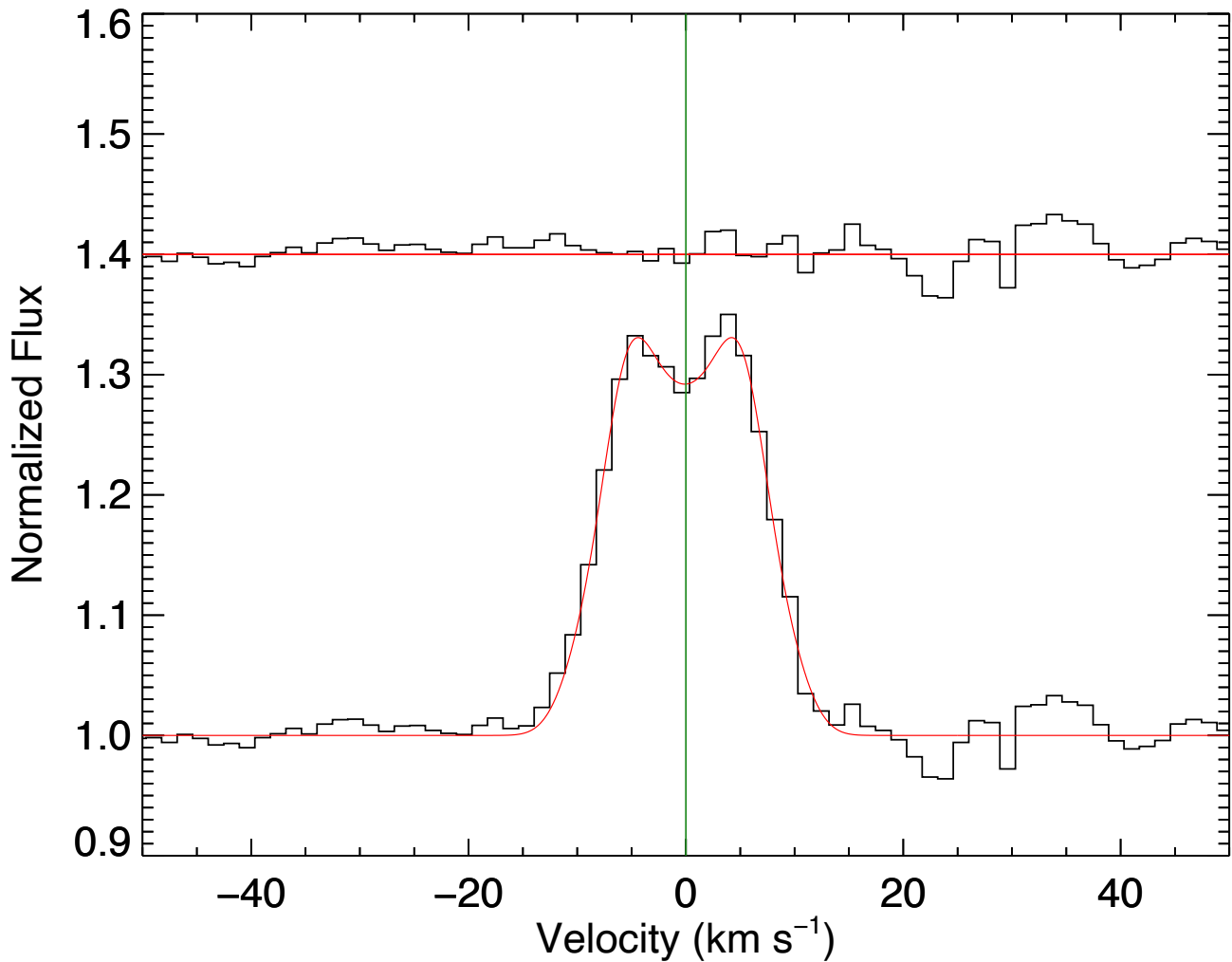}
    \caption{\label{fig:ecc_model} Eccentric disk model results for HD~141569. Below is the stacked line profile in black and model in red. Above are residuals in black and zero line in red. The eccentricity has been set to zero for this plot.}
\end{figure}

While the model itself does not capture the finer details of the physics of molecular emission, the resulting line profile is sufficient for investigating the possibility of disk eccentricity and allows for a more rapid exploration of the parameter space (hundreds of thousands or millions of runs compared to hundreds or thousands in the same time for the fluorescence model). From this we find an eccentricity upper limit of 0.02, while a sample model fit with eccentricity of 0 can be found in Figure \ref{fig:ecc_model}. The disk geometry parameters are consistent with those found using the fluorescence model.

\section{Discussion and Conclusions}
The results presented here do not confirm the possible asymmetry noted by \cite{goto06}. Our data indicates that any asymmetry at the time of
the 
observations would be less than 5\%. There are a few possibilities as to why an asymmetry may have been evident in previous data but not here. The first is that the previous asymmetry was not real. The cause may have been, for instance, artifacts in the original data. 

Alternatively, it is possible that a source of asymmetric emission may have rotated in the disk over time. An eccentric disk induced by a giant planet has the ability to create an asymmetric emission profile. Though this is not consistent with the presented data as the precession of an eccentric disk is slow, on the order of 1$^{\circ}$ per 1000 orbits \citep{kley06}. There simply is not enough time for multiple orbits between the two observations. 

If instead the original asymmetry were caused by the circumplanetary disk of a forming planet, then it would evolve on the planet's orbital timescale. At the inner rim with a 12 year difference, such a feature would move more than 100$^{\circ}$. In principle this would allow it to become hidden behind the inner rim of the circumstellar disk, similar to the case of HD~100546. However, that the gas in the disk of HD~141569 is optically thin (with emission arising from an extended, tenuous upper layer, \citealt{brittain07}) renders this unlikely as it would still be visible. 
For instance, \cite{thi14} notes that the continuum emission is optically thin in both vertical and radial directions over a range of wavelengths, to the point that the SED was not dependent on the inclination. As such, it does not seem tenable that the molecular emission from a circumplanetary disk would be hidden by the inner edge of the circumstellar disk.

An additional possibility is that the asymmetry was caused by a vortex. Vortices can be created from something as simple as perturbations within the disk (e.g. \citealt{barranco05}) or may be induced by a planet near the inner rim of the disk (e.g. \citealt{espaillat14}). Additionally, these features can be dispersed fairly rapidly due to various factors such as elliptical instability \citep{lesur09,barge16}, vertical shear \citep{barranco05}, and dust feedback \citep{fu14,crnkovic15}. The mechanism or mechanisms of dispersal will vary depending on the local conditions within the disk, as well as the vortex's radial and vertical position and structure. A deteriorating or now-destroyed vortex could result in a loss of asymmetry as observed here.
The rapidity of dissipation is a key point, and there is a wide range of lifetimes for vortices. One parameter that can affect this is the aspect ratio, a measure of elongation. For vortices with aspect ratios less than 4, dissipation can occur in less than an orbit. More elongated (aspect ratios larger than 10) can survive for hundreds of orbits. (e.g. \citealt{lesur09}). The formation of a vortex with the correct aspect ratio is a problem in and of itself. One might be created as a result of the influence of a massive planet, however it may also imply that the planet formed rapidly \citep{hammer19}.

It should be noted that a lack of direct evidence for a CPD does not imply the absence of a planet, as it may have finished accreting or entered a period of quiescence and thus be too dim to observe. A hypothetical planet may also be too small, or the surrounding disk may be too tenuous, for it to support a full CPD. Indeed, \citet{difolco20} infer that the total mass of the circumstellar disk of HD~141569 is $\rm \sim 10^{-4}M_{\odot}$. However, inferences of disk masses are highly uncertain \citep[e.g.,][]{dong2018}.

We have confirmed that the rotational temperature of CO gas is $\sim$200~K and a rotational temperature of $\sim$4500, consistent with previous results \cite{brittain02,brittain03,brittain07,thi14}. The difference in the two temperatures implies that the gas is UV fluoresced. The spectral and spectroastrometric profiles of the CO emission is consistent with Keplerian material in an axisymmetric disk. The possibility of non-axisymmetric structure is testable by additional observations at different position angles.

\section*{Acknowledgements}
This publication makes use of data products from the Wide-field Infrared Survey Explorer, which is a joint project of the University of California, Los Angeles, and the Jet Propulsion Laboratory/California Institute of Technology, funded by the National Aeronautics and Space Administration.

This work has made use of data from the European Space Agency (ESA) mission
{\it Gaia} (\url{https://www.cosmos.esa.int/gaia}), processed by the {\it Gaia}
Data Processing and Analysis Consortium (DPAC,
\url{https://www.cosmos.esa.int/web/gaia/dpac/consortium}). Funding for the DPAC
has been provided by national institutions, in particular the institutions
participating in the {\it Gaia} Multilateral Agreement.

These observations were obtained as visiting astronomers at the Infrared Telescope Facility, which is operated by the University of Hawaii under contract 80HQTR19D0030 with the National Aeronautics and Space Administration.

The authors would like to thank Jeffery Fung for his assistance and advice regarding the creation and destruction of vortices in disks.

\bibliographystyle{apj}
\bibliography{HD141569.bib}

\clearpage

\begin{table}
\begin{center}
\caption{HD~141569 observation information}
\label{tab:obstab1}
\begin{tabular}{c|c}
Date & April 19, 2017 \\
\hline
Seeing & 0.8-0.9\arcsec \\
Position angles & 356$^{\circ}$, 176$^{\circ}$ \\
Int. time (HD~141569) & 180 min\\
Int. time (HR~6556) & 6 min\\
Int. time (HR~5793) & 10 min
\end{tabular}
\end{center}
\end{table}

\clearpage

\begin{center}
\begin{longtable}{l c c}
\caption{HD~141569 CO Equivalent Width Data} \label{tab:eqw_long} \\

\multicolumn{1}{l}{{Line ID}} & \multicolumn{1}{c}{{Wavenumber}} & \multicolumn{1}{c}{{Equivalent width}} \\ 
 & (cm$^{-1}$) & (cm$^{-1}$) \\
 \vspace{1mm}\\
\hline 
\hline
 \vspace{1mm}\\
\endfirsthead

\multicolumn{3}{c}
{{\tablename\ \thetable{} -- continued}} \\

\vspace{2mm} \\

 \multicolumn{1}{l}{{Line ID}} & \multicolumn{1}{c}{{Wavenumber}} & \multicolumn{1}{c}{{Equivalent width}} \\
 \hline 
\hline
\endhead

\hline \hline
\endlastfoot

      1-0 R(12)  & 2190.02 &  $0.03049 \pm 0.00935$  \\
      1-0 R(10)  & 2183.22 &  $0.04278 \pm 0.01087$  \\
      1-0 R(8)   & 2176.28 &  $0.04682 \pm 0.00730$  \\
      1-0 R(7)   & 2172.76 &  $0.05099 \pm 0.00305$  \\
      1-0 R(6)   & 2169.20 &  $0.05845 \pm 0.00342$  \\
      1-0 R(5)   & 2165.60 &  $0.06448 \pm 0.00365$  \\
      1-0 R(4)   & 2161.97 &  $0.07791 \pm 0.00233$  \\
      1-0 R(2)   & 2154.60 &  $0.06634 \pm 0.00229$  \\
      1-0 R(1)   & 2150.86 &  $0.08847 \pm 0.00645$  \\
      1-0 P(2)   & 2135.55 &  $0.07270 \pm 0.00417$  \\
      1-0 P(5)   & 2123.70 &  $0.12240 \pm 0.00994$  \\
      1-0 P(6)   & 2119.68 &  $0.11881 \pm 0.00397$  \\
      1-0 P(7)   & 2115.63 &  $0.13156 \pm 0.00567$  \\
      1-0 P(8)   & 2111.54 &  $0.11382 \pm 0.01156$  \\
      1-0 P(11)  & 2099.08 &  $0.08902 \pm 0.01263$  \\
      1-0 P(12)  & 2094.86 &  $0.06629 \pm 0.01340$  \\
      1-0 P(13)  & 2090.61 &  $0.04340 \pm 0.01529$  \\
      1-0 P(14)  & 2086.32 &  $0.03946 \pm 0.01391$  \\
      1-0 P(17)  & 2073.26 &  $<0.00663$  \\
      1-0 P(20)  & 2059.91 &  $<0.00300$  \\
      1-0 P(21)  & 2055.40 &  $<0.00639$  \\
      1-0 P(22)  & 2050.85 &  $<0.00190$  \\
      1-0 P(23)  & 2046.28 &  $<0.00206$  \\
      1-0 P(24)  & 2041.67 &  $<0.00325$  \\
\hline
      2-1 R(20)  & 2188.49 &  $<0.00283$  \\
      2-1 R(19)  & 2185.45 &  $<0.00306$  \\
      2-1 R(18)  & 2182.36 &  $<0.00252$  \\
      2-1 R(15)  & 2172.89 &  $0.00807 \pm 0.00247$  \\
      2-1 R(14)  & 2169.66 &  $0.01839 \pm 0.00251$  \\
      2-1 R(13)  & 2166.39 &  $0.02628 \pm 0.00344$  \\
      2-1 R(12)  & 2163.08 &  $0.02909 \pm 0.00287$  \\
      2-1 R(9)   & 2152.94 &  $0.04470 \pm 0.00577$  \\
      2-1 R(8)   & 2149.49 &  $0.05101 \pm 0.02383$  \\
      2-1 R(6)   & 2142.47 &  $0.05986 \pm 0.00354$  \\
      2-1 R(5)   & 2138.91 &  $0.07514 \pm 0.01071$  \\
      2-1 R(4)   & 2135.31 &  $0.05277 \pm 0.01074$  \\
      2-1 R(2)   & 2128.01 &  $0.05586 \pm 0.01133$  \\
      2-1 R(0)   & 2120.57 &  $<0.00311$  \\
      2-1 P(1)   & 2112.98 &  $0.03020 \pm 0.00635$  \\
      2-1 P(2)   & 2109.14 &  $0.05808 \pm 0.00613$  \\
      2-1 P(3)   & 2105.26 &  $0.06359 \pm 0.01916$  \\
      2-1 P(5)   & 2097.39 &  $0.07426 \pm 0.00441$  \\
      2-1 P(6)   & 2093.41 &  $<0.00212$  \\
      2-1 P(7)   & 2089.39 &  $0.06654 \pm 0.01019$  \\
      2-1 P(11)  & 2072.99 &  $0.04170 \pm 0.01059$  \\
      2-1 P(16)  & 2051.73 &  $0.01715 \pm 0.00323$  \\
      2-1 P(20)  & 2034.14 &  $<0.00299$  \\
      2-1 P(21)  & 2029.66 &  $<0.00179$  \\
\hline
      3-2 R(22)  & 2167.25 &  $<0.00149$  \\
      3-2 R(21)  & 2164.31 &  $<0.00100$  \\
      3-2 R(20)  & 2161.34 &  $<0.00114$  \\
      3-2 R(18)  & 2155.28 &  $<0.00179$  \\
      3-2 R(17)  & 2152.20 &  $<0.00091$  \\
      3-2 R(16)  & 2149.08 &  $<0.00101$  \\
      3-2 R(14)  & 2142.72 &  $0.01384 \pm 0.00619$  \\
      3-2 R(12)  & 2136.21 &  $0.02080 \pm 0.01076$  \\
      3-2 R(11)  & 2132.91 &  $0.02213 \pm 0.00294$  \\
      3-2 R(10)  & 2129.56 &  $<0.00323$  \\
      3-2 R(9)   & 2126.18 &  $0.03317 \pm 0.00402$  \\
      3-2 R(8)   & 2122.76 &  $0.04266 \pm 0.00879$  \\
      3-2 R(7)   & 2119.30 &  $0.03226 \pm 0.00524$  \\
      3-2 R(5)   & 2112.29 &  $0.04211 \pm 0.00482$  \\
      3-2 R(4)   & 2108.72 &  $0.04307 \pm 0.01304$  \\
      3-2 R(3)   & 2105.13 &  $0.03380 \pm 0.01698$  \\
      3-2 R(1)   & 2097.82 &  $0.03703 \pm 0.00435$  \\
      3-2 R(0)   & 2094.12 &  $0.02131 \pm 0.00680$  \\
      3-2 P(1)   & 2086.60 &  $0.02656 \pm 0.00684$  \\
      3-2 P(3)   & 2078.95 &  $0.03738 \pm 0.02623$  \\
      3-2 P(5)   & 2071.15 &  $0.05154 \pm 0.00548$  \\
      3-2 P(6)   & 2067.21 &  $0.05525 \pm 0.01392$  \\
      3-2 P(9)   & 2055.16 &  $0.03221 \pm 0.01300$  \\
      3-2 P(10)  & 2051.08 &  $0.03998 \pm 0.00820$  \\
      3-2 P(11)  & 2046.96 &  $0.03404 \pm 0.00875$  \\
      3-2 P(14)  & 2034.41 &  $0.01792 \pm 0.01090$  \\
      3-2 P(15)  & 2030.16 &  $0.00704 \pm 0.00685$  \\
      3-2 P(17)  & 2021.56 &  $<0.00187$  \\
      3-2 P(20)  & 2008.42 &  $<0.00161$  \\
      3-2 P(21)  & 2003.98 &  $<0.00481$  \\
      3-2 P(22)  & 1999.50 &  $<0.00217$  \\
      3-2 P(23)  & 1994.99 &  $<0.00198$  \\
      3-2 P(24)  & 1990.45 &  $<0.00264$  \\
\hline
      4-3 R(21)  & 2137.20 &  $<0.00690$  \\
      4-3 R(19)  & 2131.29 &  $<0.00229$  \\
      4-3 R(18)  & 2128.27 &  $<0.00782$  \\
      4-3 R(17)  & 2125.22 &  $<0.00177$  \\
      4-3 R(16)  & 2122.14 &  $<0.00856$  \\
      4-3 R(15)  & 2119.01 &  $<0.00167$  \\
      4-3 R(13)  & 2112.65 &  $0.00819 \pm 0.00366$  \\
      4-3 R(12)  & 2109.41 &  $0.01689 \pm 0.00627$  \\
      4-3 R(11)  & 2106.14 &  $<0.00183$  \\
      4-3 R(10)  & 2102.83 &  $<0.01172$  \\
      4-3 R(9)   & 2099.48 &  $0.02340 \pm 0.00849$  \\
      4-3 R(8)   & 2096.10 &  $0.02407 \pm 0.00653$  \\
      4-3 R(7)   & 2092.68 &  $0.02295 \pm 0.00599$  \\
      4-3 R(6)   & 2089.22 &  $0.01877 \pm 0.00679$  \\
      4-3 R(5)   & 2085.73 &  $0.02908 \pm 0.00608$  \\
      4-3 R(3)   & 2078.64 &  $<0.01046$  \\
      4-3 R(1)   & 2071.40 &  $0.02445 \pm 0.00407$  \\
      4-3 R(0)   & 2067.74 &  $0.01236 \pm 0.01187$  \\
      4-3 P(3)   & 2052.71 &  $0.03673 \pm 0.01844$  \\
      4-3 P(7)   & 2037.12 &  $0.03544 \pm 0.01078$  \\
      4-3 P(8)   & 2033.14 &  $0.03008 \pm 0.00774$  \\
      4-3 P(9)   & 2029.13 &  $0.02965 \pm 0.00634$  \\
      4-3 P(11)  & 2021.00 &  $0.01574 \pm 0.00808$  \\
      4-3 P(15)  & 2004.34 &  $<0.00172$  \\
      4-3 P(16)  & 2000.09 &  $<0.00302$  \\
      4-3 P(17)  & 1995.81 &  $<0.00159$  \\
      4-3 P(20)  & 1982.78 &  $<0.00202$  \\
      4-3 P(21)  & 1978.37 &  $<0.00151$  \\
      4-3 P(22)  & 1973.93 &  $<0.00158$  \\
      4-3 P(23)  & 1969.45 &  $<0.00317$  \\
      4-3 P(24)  & 1964.95 &  $<0.00350$  \\
\hline
      5-4 R(21)  & 2110.15 &  $<0.00167$  \\
      5-4 R(19)  & 2104.31 &  $<0.00165$  \\
      5-4 R(17)  & 2098.32 &  $<0.00245$  \\
      5-4 R(16)  & 2095.26 &  $<0.00266$  \\
      5-4 R(15)  & 2092.17 &  $0.00869 \pm 0.00847$  \\
      5-4 R(14)  & 2089.05 &  $<0.00146$  \\
      5-4 R(13)  & 2085.88 &  $<0.00379$  \\
      5-4 R(11)  & 2079.44 &  $0.00759 \pm 0.00550$  \\
      5-4 R(10)  & 2076.17 &  $<0.00196$  \\
      5-4 R(9)   & 2072.85 &  $0.01265 \pm 0.01087$  \\
      5-4 R(8)   & 2069.51 &  $0.01840 \pm 0.01024$  \\
      5-4 R(7)   & 2066.12 &  $0.01955 \pm 0.01380$  \\
      5-4 R(4)   & 2055.75 &  $0.02450 \pm 0.00645$  \\
      5-4 R(3)   & 2052.22 &  $0.04551 \pm 0.01377$  \\
      5-4 R(2)   & 2048.66 &  $0.02031 \pm 0.00439$  \\
      5-4 P(1)   & 2034.05 &  $0.00982 \pm 0.00519$  \\
      5-4 P(2)   & 2030.31 &  $0.01924 \pm 0.00980$  \\
      5-4 P(6)   & 2015.00 &  $0.02174 \pm 0.01102$  \\
      5-4 P(7)   & 2011.09 &  $0.02277 \pm 0.00494$  \\
      5-4 P(8)   & 2007.14 &  $0.02341 \pm 0.00958$  \\
      5-4 P(9)   & 2003.17 &  $<0.00253$  \\
      5-4 P(10)  & 1999.15 &  $<0.00682$  \\
      5-4 P(11)  & 1995.10 &  $0.01221 \pm 0.00753$  \\
      5-4 P(14)  & 1982.76 &  $<0.00183$  \\
      5-4 P(15)  & 1978.58 &  $<0.00156$  \\
      5-4 P(16)  & 1974.37 &  $0.01206 \pm 0.00333$  \\
      5-4 P(17)  & 1970.13 &  $<0.00226$  \\
      5-4 P(19)  & 1961.54 &  $<0.00628$  \\
      5-4 P(20)  & 1957.20 &  $<0.00242$  \\
      5-4 P(21)  & 1952.82 &  $<0.00200$  \\
\hline
      6-5 R(22)  & 2086.00 &  $<0.00393$  \\
      6-5 R(20)  & 2080.31 &  $<0.00200$  \\
      6-5 R(18)  & 2074.46 &  $<0.00298$  \\
      6-5 R(17)  & 2071.48 &  $<0.00397$  \\
      6-5 R(16)  & 2068.46 &  $<0.00157$  \\
      6-5 R(15)  & 2065.41 &  $<0.00373$  \\
      6-5 R(14)  & 2062.32 &  $<0.00311$  \\
      6-5 R(11)  & 2052.81 &  $<0.00392$  \\
      6-5 R(10)  & 2049.57 &  $0.01325 \pm 0.00558$  \\
      6-5 R(8)   & 2042.98 &  $0.00982 \pm 0.00797$  \\
      6-5 R(7)   & 2039.63 &  $0.01580 \pm 0.00971$  \\
      6-5 R(6)   & 2036.25 &  $0.01719 \pm 0.00582$  \\
      6-5 R(2)   & 2022.35 &  $0.01539 \pm 0.01396$  \\
      6-5 R(0)   & 2015.18 &  $<0.00289$  \\
      6-5 P(2)   & 2004.17 &  $<0.00595$  \\
      6-5 P(3)   & 2000.43 &  $0.01395 \pm 0.00848$  \\
      6-5 P(4)   & 1996.66 &  $0.01938 \pm 0.00988$  \\
      6-5 P(6)   & 1989.01 &  $0.01743 \pm 0.01661$  \\
      6-5 P(9)   & 1977.27 &  $0.01457 \pm 0.01252$  \\
      6-5 P(10)  & 1973.29 &  $<0.00118$  \\
      6-5 P(11)  & 1969.28 &  $<0.00341$  \\
      6-5 P(12)  & 1965.24 &  $<0.00441$  \\
      6-5 P(14)  & 1957.05 &  $<0.00465$  \\
      6-5 P(15)  & 1952.90 &  $<0.00150$  \\
\hline
       7-6 R(22) & 2059.06 &  $<0.00144$  \\
       7-6 R(21) & 2056.27 &  $<0.00465$  \\
       7-6 R(19) & 2050.57 &  $<0.00412$  \\
       7-6 R(18) & 2047.66 &  $<0.00261$  \\
       7-6 R(16) & 2041.73 &  $<0.00516$  \\
       7-6 R(15) & 2038.71 &  $<0.00393$  \\
       7-6 R(14) & 2035.66 &  $<0.00101$  \\
       7-6 R(13) & 2032.56 &  $<0.00106$  \\
       7-6 R(11) & 2026.26 &  $<0.00253$  \\
        7-6 R(9) & 2019.81 &  $<0.00227$  \\
        7-6 R(7) & 2013.22 &  $0.01040 \pm 0.00949$  \\
        7-6 R(6) & 2009.87 &  $0.00843 \pm 0.00777$  \\
        7-6 R(4) & 2003.06 &  $0.03692 \pm 0.03516$  \\
        7-6 R(3) & 1999.60 &  $<0.00248$  \\
        7-6 R(2) & 1996.11 &  $0.00499 \pm 0.00488$  \\
        7-6 R(0) & 1989.01 &  $<0.00632$  \\
        7-6 P(1) & 1981.78 &  $<0.00238$  \\
        7-6 P(2) & 1978.11 &  $<0.00162$  \\
        7-6 P(3) & 1974.40 &  $<0.00119$  \\
        7-6 P(4) & 1970.66 &  $<0.00209$  \\
        7-6 P(6) & 1963.08 &  $0.02082 \pm 0.00560$  \\
        7-6 P(7) & 1959.24 &  $<0.00154$  \\
        7-6 P(8) & 1955.36 &  $<0.00678$  \\
        7-6 P(9) & 1951.45 &  $<0.00269$  \\
\hline
  (13)1-0 R(22)  & 2171.31 &  $<0.00230$  \\
  (13)1-0 R(21)  & 2168.42 &  $<0.00128$  \\
  (13)1-0 R(19)  & 2162.54 &  $<0.00105$  \\
  (13)1-0 R(17)  & 2156.51 &  $<0.00431$  \\
  (13)1-0 R(16)  & 2153.44 &  $<0.00129$  \\
  (13)1-0 R(15)  & 2150.34 &  $<0.00179$  \\
  (13)1-0 R(13)  & 2144.03 &  $0.00831 \pm 0.00450$  \\
  (13)1-0 R(11)  & 2137.59 &  $0.01118 \pm 0.00536$  \\
  (13)1-0 R(9)   & 2131.00 &  $0.01725 \pm 0.00391$  \\
  (13)1-0 R(5)   & 2117.43 &  $0.02828 \pm 0.00892$  \\
  (13)1-0 R(4)   & 2113.95 &  $0.02591 \pm 0.00348$  \\
  (13)1-0 R(3)   & 2110.44 &  $0.01988 \pm 0.00258$  \\
  (13)1-0 R(2)   & 2106.90 &  $0.02079 \pm 0.00745$  \\
  (13)1-0 R(0)   & 2099.71 &  $0.01035 \pm 0.00993$  \\
  (13)1-0 P(1)   & 2092.39 &  $0.01080 \pm 0.00937$  \\
  (13)1-0 P(2)   & 2088.68 &  $0.02318 \pm 0.00828$  \\
  (13)1-0 P(7)   & 2069.66 &  $0.03204 \pm 0.00815$  \\
  (13)1-0 P(9)   & 2061.82 &  $0.03615 \pm 0.00746$  \\
  (13)1-0 P(10)  & 2057.86 &  $0.02184 \pm 0.00709$  \\
  (13)1-0 P(11)  & 2053.86 &  $0.02056 \pm 0.00546$  \\
  (13)1-0 P(12)  & 2049.83 &  $0.01428 \pm 0.00528$  \\
  (13)1-0 P(14)  & 2041.69 &  $<0.00597$  \\
  (13)1-0 P(15)  & 2037.57 &  $<0.00411$  \\
  (13)1-0 P(16)  & 2033.42 &  $<0.00363$  \\
  (13)1-0 P(17)  & 2029.24 &  $<0.00387$  \\
  (13)1-0 P(19)  & 2020.78 &  $<0.00198$  \\
  (13)1-0 P(21)  & 2012.21 &  $<0.00121$  \\
  (13)1-0 P(22)  & 2007.88 &  $<0.00467$  \\
  (13)1-0 P(23)  & 2003.51 &  $<0.00313$  \\
  (13)1-0 P(24)  & 1999.12 &  $<0.00832$  \\
\hline
  (13)2-1 R(21)  & 2142.39 &  $<0.00117$  \\
  (13)2-1 R(19)  & 2136.57 &  $<0.00203$  \\
  (13)2-1 R(18)  & 2133.60 &  $<0.00094$  \\
  (13)2-1 R(17)  & 2130.61 &  $<0.00108$  \\
  (13)2-1 R(15)  & 2124.50 &  $<0.01001$  \\
  (13)2-1 R(13)  & 2118.26 &  $<0.00187$  \\
  (13)2-1 R(12)  & 2115.09 &  $<0.00349$  \\
  (13)2-1 R(11)  & 2111.88 &  $<0.00143$  \\
  (13)2-1 R(9)   & 2105.36 &  $<0.00874$  \\
  (13)2-1 R(7)   & 2098.71 &  $0.01535 \pm 0.00418$  \\
  (13)2-1 R(6)   & 2095.33 &  $0.01329 \pm 0.00491$  \\
  (13)2-1 R(5)   & 2091.92 &  $0.01841 \pm 0.00330$  \\
  (13)2-1 R(2)   & 2081.49 &  $0.01540 \pm 0.01090$  \\
  (13)2-1 R(0)   & 2074.36 &  $0.00770 \pm 0.00705$  \\
  (13)2-1 P(6)   & 2048.41 &  $0.02421 \pm 0.00403$  \\
  (13)2-1 P(8)   & 2040.70 &  $0.02197 \pm 0.01111$  \\
  (13)2-1 P(9)   & 2036.80 &  $0.01567 \pm 0.00570$  \\
  (13)2-1 P(11)  & 2028.91 &  $0.01185 \pm 0.00624$  \\
  (13)2-1 P(13)  & 2020.89 &  $<0.00236$  \\
  (13)2-1 P(18)  & 2000.30 &  $<0.00189$  \\
  (13)2-1 P(19)  & 1996.09 &  $0.01091 \pm 0.00787$  \\
  (13)2-1 P(21)  & 1987.58 &  $<0.00724$  \\
  (13)2-1 P(22)  & 1983.28 &  $<0.00227$  \\
  (13)2-1 P(23)  & 1978.95 &  $<0.00162$  \\
  (13)2-1 P(24)  & 1974.59 &  $<0.00090$  
\end{longtable}
\end{center}

\begin{table}
\begin{center}
\caption{Rotational tempeatures}
\label{tab:rottemp}
\begin{tabular}{cccc}
Ro-vibrational & Temperature \\
band & (K) \\
\hline
\hline
$^{12}$CO v=1-0 & 195$\pm$25 \\
$^{12}$CO v=2-1 & 217$\pm$13 \\
$^{12}$CO v=3-2 & 193$\pm$16 \\
$^{12}$CO v=4-3 & 196$\pm$30 \\
$^{12}$CO v=5-4 & 251$\pm$47 \\
$^{12}$CO v=6-5 & 219$\pm$116 \\
$^{12}$CO v=7-6 & 149$\pm$587\\
\hline
$^{13}$CO v=1-0 & 209$\pm$28 \\
$^{13}$CO v=2-1 & 144$\pm$42 \\
\hline
$^{12}$CO Vibrational & 4537$\pm$436 
\end{tabular}
\end{center}
\end{table}

\begin{table}
\begin{center}
\caption{Final model parameters}
\label{tab:finalparam}
\begin{tabular}{l r}
Parameter & Value \\
\hline
\hline
Fixed \\
\hline
M$_\star$ & 2.39 M$_\odot$ \\
L$_{rel}$ & 1.2 \\
$i$ & 51$^{\circ}$ \\
$v_{obs}$ & -21.4 km/s \\
\hline
Varied \\
\hline
R$_{in}$ & 13$\pm$2 au \\
R$_{out}$ & 60$\pm$5 au \\
T$_{rot,0}$ & 495$\pm$5 K \\
$\alpha _{rot}$ & 0.275$\pm$0.020 \\
v$_{turb}$ & 2.5$\pm$ 0.5 km/s \\
Flare angle & $16^{\circ}$$^{+13^{\circ}}_{-15^{\circ}}$

\end{tabular}
\end{center}
\end{table}

\end{document}